\documentclass[twocolumn,useAMS,usenatbib]{mn2e}
\usepackage{amssymb}
\usepackage[pdftex]{graphicx}

\newcommand{\Ms}{\ensuremath{M_{\odot}}}
\newcommand{\eg}{{\it e.g.}}
\newcommand{\cf}{{\it c.f.~}}
\newcommand{\ie}{{\it i.e.}}
\newcommand{\viz}{{\it viz.}}
\newcommand{\mnras}{{\it MNRAS}}
\newcommand{\aap}{{\it A\&A}}
\newcommand{\apj}{{\it ApJ}}
\newcommand{\apjl}{{\it ApJ}}
\newcommand{\apjs}{{\it ApJS}}
\newcommand{\aj}{{\it AJ}}
\newcommand{\araa}{{\it ARA\&A}}
\title[Formation of NGC 3603 young cluster]
{The formation of NGC 3603 young starburst cluster: ``prompt'' hierarchical assembly or
monolithic starburst?}
\author[Banerjee \& Kroupa]
{Sambaran Banerjee$^1$\thanks{Corresponding author. E-mail: sambaran@astro.uni-bonn.de}
and Pavel Kroupa$^1$\\
$^1$University of Bonn, Auf dem H\"ugel 71, D-53121, Bonn, Germany}
\begin{document}

\date{Submitted August, 2014}


\maketitle

\label{firstpage}

\begin{abstract}
The formation of very young massive clusters or ``starburst'' clusters is currently
one of the most widely debated topic in astronomy. The classical notion dictates
that a star cluster is formed in-situ in a dense molecular gas clump. The stellar radiative
and mechanical feedback to the residual gas energizes the latter until it escapes the system.
The newly born gas-free young cluster eventually readjusts with the corresponding mass loss.
Based on the observed substructured morphologies of many young stellar associations,
it is alternatively suggested that even the smooth-profiled massive clusters are
also assembled from migrating less massive subclusters. A very young (age $\approx$ 1 Myr),
massive ($>10^4\Ms$) star cluster like the Galactic NGC 3603 young cluster (HD 97950) is
an appropriate testbed for distinguishing between the above ``monolithic'' and
``hierarchical'' formation scenarios. A recent study by Banerjee and Kroupa (2014)
demonstrates that the monolithic scenario remarkably reproduces the HD 97950 cluster. In particular,
its shape, internal motion and the mass distribution of stars are found to follow naturally
and consistently from a single model calculation undergoing $\approx$70\% by mass gas dispersal.
In the present work, we explore the possibility of the formation of the above cluster via hierarchical 
assembly of subclusters. These subclusters are initially distributed over a wide range of spatial volumes
and have various modes of sub-clustering in both absence and presence of a background gas potential.
Unlike the above monolithic initial system that reproduces HD 97950 very well, the same is found to be
prohibitive with hierarchical assembly alone (with/without a gas potential).
Only those systems which assemble promptly into a single cluster (in $\lesssim1$ Myr) from a close separation
(all within $\lesssim2$ pc) could match the observed density profile of HD 97950 after a
similar gas removal. These results therefore suggest that the NGC 3603 young cluster has formed 
essentially monolithically, \ie, either in-situ or via a prompt assembly, followed by a substantial residual
gas expulsion. Both scenarios are consistent with the inferred young age
and the small age spread of this cluster.
Future observations of molecular cloud filaments with \emph{ALMA} and proper motion measurements of young clusters
with \emph{Gaia} will provide more direct tests of such birth environments.
\end{abstract}


\section{Introduction}\label{intro}

How very young (age of a few Myr), massive (rich) star clusters (hereafter VYMCs as coined by \citealt{sb2014}; also
known as ``starburst'' clusters) of near-spherical shape form out of vast molecular clouds is
one of the most debated topic in astrophysics. The most massive ones are typically found
in an overall near-spherical core-halo form and they can be surrounded by a cocoon of
HII (ionized hydrogen) region. Perhaps the most widely
discussed among such systems are the R136 cluster in the Large Magellanic Cloud (LMC), the central cluster
(HD 97950; hereafter HD97950) of the Galactic ``Giant Nebula'' NGC 3603 and the Arches cluster \citep{pz2010}. Several VYMCs
are found as the richest member of extended cluster complexes/stellar associations, \eg, the ONC \citep{ab2012}.
Young stellar systems are also found as extended associations of OB stars, \eg, the Cygnus OB2 \citep{kuhn2014,wrt2014}.
The highly compact young cluster RCW 38 is a special case which is mostly embedded in molecular
hydrogen except its central region \citep{derose2009}. In several cases there are indications that even
an overall core-halo shaped VYMC also contains substructures,
as obtained from the ``Massive Young Star-Forming Complex Study
in Infrared and X-ray'' (MYStIX; \citealt{fglsn2013})
survey catalog \citep{kuhn2014}.

At present, there exist apparently at least two distinct scenarios for formation of VYMCs. The monolithic or
episodic (top-down) scenario dictates the formation of a compact star cluster in an
essentially single but highly active star-formation episode. The infant cluster of pre-main-sequence (PMS) and
main sequence (MS) stars remains embedded in its parent molecular gas cloud where the latter gets energized
(and ionized) by the radiation and material outflows from the stars. Such energy injection eventually causes the embedding gas
to become gravitationally unbound from the system and disperse in a timescale typically comparable
to the dynamical time of the stellar system, \ie, too fast for the stars to adjust with the 
corresponding reduction of potential well. This causes the gas-free stellar system to expand violently and lose
a fraction of its stars depending on its initial mass and concentration
\citep{adm2000,pketl2001,bk2007,sb2013,pfkz2013}. The remaining system
may eventually regain virial or dynamical equilibrium (re-virialization); hence a particular VYMC may or may not be in
dynamical equilibrium depending on the time taken to re-virialize and
the epoch at which it is observed \citep{sb2013}. Such a ``monolithic'' or top-down scenario has successfully explained
well observed VYMCs, \eg, ONC (also the Pleiades; \citealt{pketl2001}), R136 \citep{sb2013}
and the NGC 3603 young cluster \citep{sb2014}.

Alternatively, VYMCs are thought to have formed ``bottom-up'' via merging of less massive subclusters \citep{longm2014}.
These subclusters are usually thought to form with high local star formation efficiency (SFE),
\ie, they convert a large fraction of the gas in their vicinity into stars. Several of such subclusters fall
onto each other
and merge to form the final VYMC. The gravitational potential of the background molecular gas within which
these subclusters appear augments the infall rate
(the so-called ``conveyer belt mechanism''; \citealt{longm2014}). The observational motivation for such a scenario is the apparent
substructures in OB associations and even in VYMCs having overall core-halo configurations \citep{kuhn2014}.
On the theoretical side, star formation has been investigated in hydrodynamic calculations involving
development of seed turbulence, in cubical/spherical gas clouds, into high-density filaments where star (sink particle) formation
occurs as a result of gravitational collapse and fragmentation \citep{kl1998,bate2004,giri2011}.

In such calculations, small clusters of
proto-stars are formed in filaments and/or filament nodes over a short period of time,
which then fall collectively into the potential of the cloud to form larger (gas-embedded) clusters
(\eg, in \citealt{bate2009} and \citealt{giri2011}).
Different groups have reached the state-of-the-art of such calculations by incorporating physical processes
in different degrees of detail but for mass scales much lighter than
VYMCs. Such smoothed-particle-hydrodynamic (SPH) self-star-forming calculations,
requiring very high particle resolution, is formidable for the mass range of VYMCs ($>10^4\Ms$). 
High-resolution (reaching the ``opacity limit'') SPH computations
have so far been done forming stars in spherical gas clouds of up to
$\approx500\Ms$ only \citep{kl1998,bate2004,bate2009,giri2011,giri2012,bate2012}
but without any feedback and hence self-regulation mechanism. Radiation-magnetohydrodynamic (MHD) calculations
including stellar feedback (radiation and matter outflows)
to the star-forming gas have also been carried out from proto-stellar scales
\citep{mnm2012,bate2013} up to $\approx50\Ms$ gas spheres \citep{PriceBate2010}.
While the latter studies provide insights into the self-regulation mechanisms in the star formation process
and point to an SFE near $30$\%, consistently with observations \citep{lnl2003},
the processes that lead to the ultimate dispersal of the residual gas are still unclear 
(Matthew Bate, University of Exeter, U.K.: private communication).
More recently, an independent semi-analytical
study \citep{rb2014} of formation of clump-cores (that would eventually turn into proto-stars)
in gas clumps and of the maximum mass of the cores infers an upper limit
of $\approx30$\% for the clump SFE. This is consistent with the hydrodynamic calculations
with self-regulation and observations in the solar neighborhood (see above).

Note that
the gas has to disperse from the region in the molecular cloud where the cluster ultimately assembles,
to obtain a gas-free young cluster like what we see today. The only way the essential dynamical effects
of gas dispersal can be included is to adopt a time-varying background potential mimicking the effect of the gas,
which is widely used \citep{pketl2001,pfkz2013,sb2014}.
The latter approach has successfully explained several well observed VYMCs, \viz, the Galactic ONC \citep{pketl2001}
and NGC 3603 \citep{sb2014} and R136 \citep{sb2013} of the LMC. These studies point to a universal SFE of
$\epsilon\approx33$\% and a sonic dispersal of the residual ionized hydrogen or HII gas 
(see \citealt{sb2013} and references therein), remarkably reproducing the measured kinematic
and structural properties of these clusters. The dynamical process of coalescence
of subclusters into more massive clusters has also been studied recently using direct
N-body calculations in both absence \citep{fuji2012} and presence \citep{sm2013} of
a background gas potential. The role of this process is also investigated in the context
of formation of dwarf galaxies through merger of young massive clusters \citep{ams2014}.

\begin{figure*}
\centering
\includegraphics[width=14cm, angle=0]{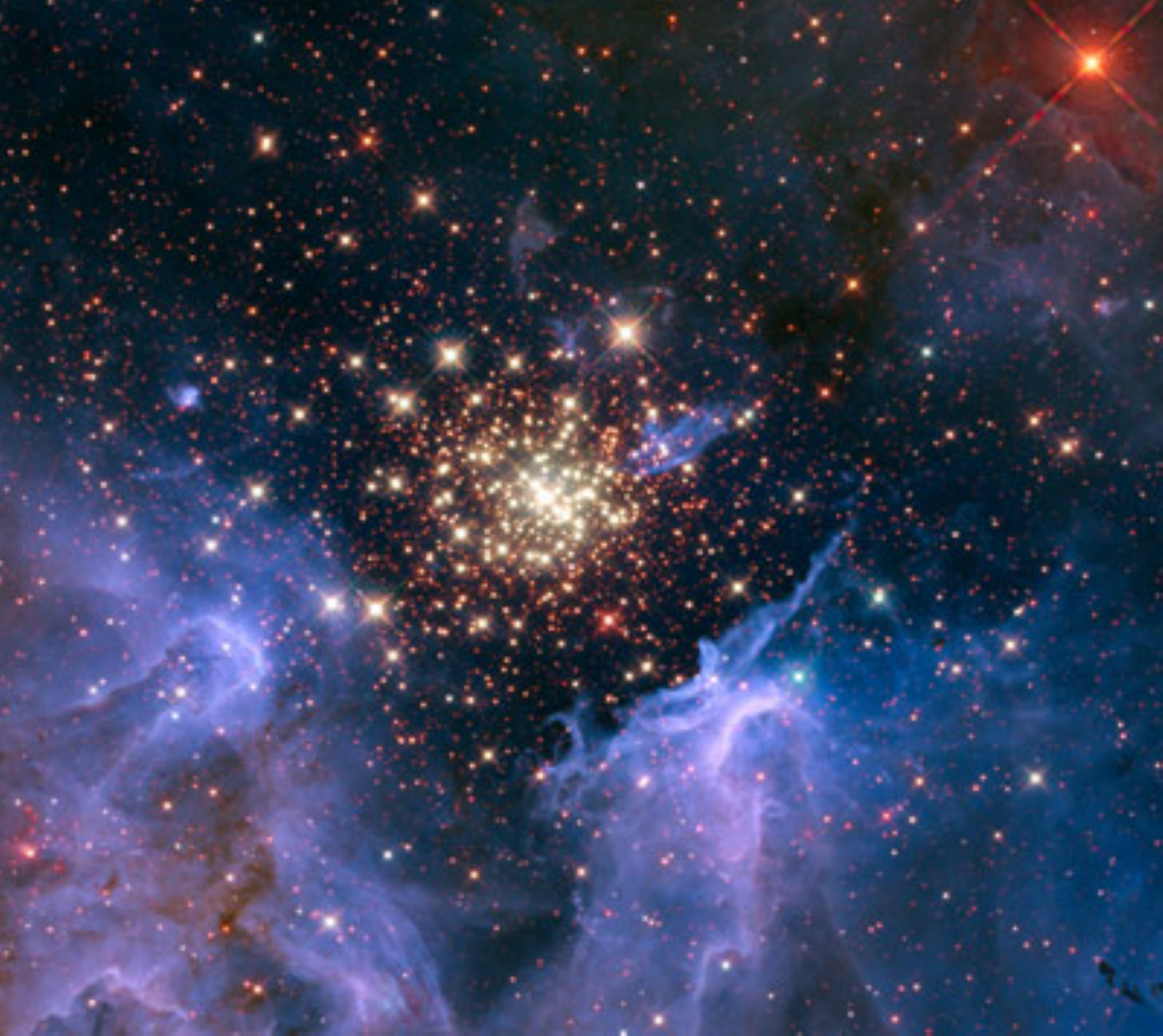}
\caption{An image of NGC 3603 where the central star cluster is HD 97950 which we
aim to reproduce, from model N-body calculations, in this study. This is a
composite image obtained from the Hubble Space Telescope using the filters
F128N (Paschen-beta), F164N ($\left[{\rm Fe~II}\right]$), F555W ($V$),
F657N (H-alpha) and F673N ($\left[{\rm S~II}\right]$). The image is
$\approx180''$ (5 pc) wide.
\emph{Credit:}  NASA, ESA, R. O'Connell (University of Virginia),
F. Paresce (National Institute for Astrophysics, Bologna, Italy), 
E. Young (Universities Space Research Association/Ames Research Center),
the WFC3 Science Oversight Committee, and the Hubble Heritage Team (STScI/AURA).
This image is obtained from the online resource \emph{hubblesite.org} (public domain).}
\label{fig:ngc3603_HST}
\end{figure*}

The work reported in the present paper focuses on formation of VYMCs through subcluster
merging. In particular, we consider the reproducibility of the VYMC HD97950 in the Galactic
NGC 3603 star forming region from a wide range of initially subclustered conditions. As in \citet{sb2014},
we target this cluster again due to the availability of its detailed, high-quality
structural and kinematical data
from the Very Large Telescope (VLT; \citealt{hara2008}) and the Hubble Space Telescope
(HST; \citealt{roch2010,pang2013}). Situated on the Carina spiral arm of the Milky way at
a distance of $\approx6-7$ kpc from the Sun, HD97950 is our nearest starburst
cluster and also perhaps the most well observed one. Furthermore, its large
mass (photometrically $10000\Ms-16000\Ms$; \citealt{stol2006,hara2008}), extreme young
age of $\approx1$ Myr with a small age spread \citep{stol2004,pang2013} despite
having a clear core-halo configuration makes HD97950 ideal for testing theories of VYMC/globular
cluster formation. Fig.~\ref{fig:ngc3603_HST} shows an HST image of HD97950.
Table~\ref{tab:hd9795} summarizes the observed properties of HD97950 which
will be referred to in the following sections at different occasions.

We explore a wide range of initial subclustering mode and the spatial volume over
which the subclusters are distributed. We also explore the effects of the presence
of a background gas potential.
We demonstrate that prompt merging of subclusters that fall onto each other from close
initial separations ($\lesssim1{\rm~pc}$) followed by a substantial gas expulsion ($\approx 70$\%
by mass) can reproduce the observed surface density profile of the HD97950 cluster.
These calculations also suggest that a significant amount of gas expulsion is essential for getting the observed
profile right. Nevertheless, the overall agreement with both the structural and kinematic data
of the cluster is better achieved with the monolithic model of \citet{sb2014} than the hierarchical
formation models as computed here.

\section{Initial conditions}\label{initcond}

In the following sections we detail our calculations that aim to form the observed
HD97950 cluster through subcluster merging, starting from a range of initial configurations.
We also discriminate the cases where the subclusters fall in under the influence of a
background gas potential and those where the infall occurs only under the mutual gravity of the
subclusters. The objective is to determine the timescales for formation of core-halo configurations
beginning from various substructured initial conditions vis-a-vis the present age of HD97950.
We also determine the conditions under which the final core-halo cluster agrees with the observed
profile and kinematics of HD97950.

\subsection{Initial (proto) stellar distribution}\label{initstar}

The substructured initial conditions are generated by distributing compact Plummer spheres 
\citep{plum1911} uniformly
over a spherical volume of radius $R_0$. The total stellar mass distributed in this way
is always the lower photometric mass estimate of $M_\ast\approx10^4\Ms$ for HD97950. One motivation
for adopting the lower mass limit is the study of \citet{sb2014} where a monolithic
embedded cluster of the above total stellar mass turns out to be optimal in reproducing the HD97950 cluster
after gas expulsion. As we will see below, this is true in this case also.

The above fashion of initial subclustering is an idealization and extrapolation of what is found in the
largest SPH calculations of cluster formation to date \citep{bate2009,bate2012,giri2011}.
In these calculations, the uniformly dense and turbulent gas spheres develop multiple
high-density filamentary structures (as what is observed in molecular clouds)
where typically 1-2 subclusters appear per filament
at their densest points and also at the filament junctions. This
is consistent with what is observed in dense molecular gas filaments, \eg, through the
\emph{Herschel} telescope \citep{schn2010,schn2012}.
These filaments often extend
across the spherical cloud and can have arbitrarily bent configuration so that subclusters can
appear anywhere within the sphere. Also, most of the filaments and the subclusters are
found to form promptly. Hence, although the total gas mass used in these SPH calculations are
about an order of magnitude smaller than the above $M_\ast$ and the total mass of
proto-stars formed is up to 20\% of the gas mass \citep{giri2011}, the above mode
of subclustering is an idealized but appropriate extension of the state-of-the-art
hydrodynamic calculations.

The Plummer shape of the subclusters also conform with the observed Plummer
profile of the filament sections in molecular clouds \citep{mali2012}. We choose   
the half-mass radii, $r_h(0)$, of these Plummer spheres to be typically
between $0.1-0.3$ pc, in accordance with the observed widths of these
highly compact molecular-cloud filaments \citep{andr2011,schn2012}. Such
compact sizes of the subclusters are also consistent with those observed in stellar complexes, \eg,
the Taurus-Auriga stellar groups \citep{palsth2002,kb2003}. However, in some
calculations, we also use larger $r_h(0)$ (see below).

\begin{figure*}
\centering
\includegraphics[width=14.0cm, angle=0]{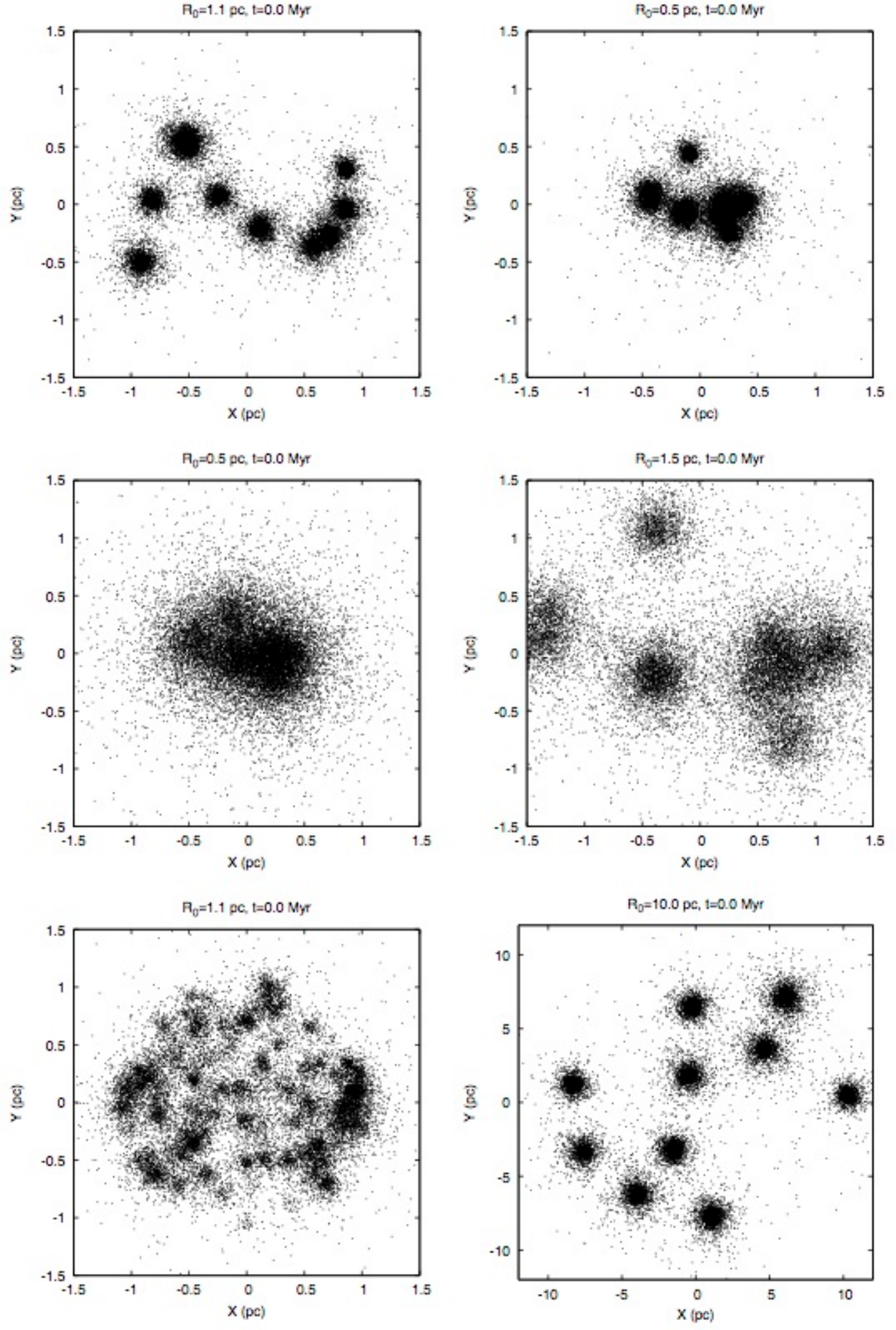}
\caption{The primary varieties of the \emph{initial} configurations that are evolved in this study,
shown in projection. In this and all the subsequent figures, the panels are numbered
left-to-right, top-to-bottom.
In each case, a set of Plummer spheres (subclusters) are uniformly distributed over a spherical
volume of radius $R_0$, totalling a stellar mass of
$M_\ast\approx10000\Ms$. Panels 1, 2, 3, 4 and 6 are examples of type A or
``blobby'' systems containing 10 subclusters of $m_{cl}(0)\approx10^3\Ms$ each.
With smaller $R_0$, the subclusters overlap more with each other (\cf, panels 1 \& 2
with subcluster half mass radius $r_h(0)\approx0.1$ pc and panels
3 \& 4 with $r_h(0)\approx0.3$ pc). This is also true for increasing $r_h(0)$
(\cf, panels 1 \& 4). Panel 5 is an
example of type B or ``grainy'' initial configuration containing $\approx 150$
subclusters of mass range $10\Ms\lesssim m_{cl}(0) \lesssim 100\Ms$.
While panels 1-5 are examples of ``compact'' configurations,
for which $R_0\leq2.5$ pc, panel 6, with $R_0=10$ pc, represents an ``extended''     
configuration where the subclusters are much more distinct. 
See Sec.~\ref{initcond} for details of the initial setups.
Tables~\ref{tab:morphevol} and \ref{tab:kfit1} provide a complete list
of initial systems computed here.
}
\label{fig:snaps_0myr}
\end{figure*}

The number of subclusters, $n$, over which the $M_\ast\approx10^4\Ms$ is subdivided 
has to be chosen somewhat arbitrarily (as in recent studies, \eg, \citealt{fuji2012}).
In the above mentioned hydrodynamic calculations, subclusters containing 10s to 100s of proto-stars
(sink particles) could form (subject to the caveat that sink particle formation continues
unhindered by stellar feedback).
However, it is not straightforward to extrapolate the richness of
subclusters to the much larger masses as in here. In order to incorporate a range of possibilities,
we consider two primary cases of the initial subdivision of the total stellar mass.
The ``blobby'' (type A) systems consist of 10 subclusters 
of $m_{cl}(0)\approx10^3\Ms$ each. Panels 1, 2, 4 and 6 of Fig.~\ref{fig:snaps_0myr}
are examples of such initial systems. Note that in
this and all the subsequent figures, the panels are numbered
left-to-right, top-to-bottom.  On the other hand, ``grainy'' (type B) systems comprise
$\approx 150$ subclusters with mass range $10\Ms\lesssim m_{cl}(0) \lesssim 100\Ms$
summing up to $M_\ast\approx10^4\Ms$. We show below that the mode of initial subdivision
does not influence the primary results.

The initial spanning radius, $R_0$, is chosen over a wide range, \viz, $0.5{\rm~pc}\lesssim R_0\lesssim 10.0{\rm~pc}$,
to explore the wide range of molecular cloud densities (see below) and spatial extents as observed in star-forming regions
and stellar complexes. Tables~\ref{tab:morphevol} and \ref{tab:kfit1} (their first and second columns)
provide a complete list of the initial conditions for the computations here. The detailed nomenclature, in their
first columns, is explained in Table~\ref{tab:morphevol} and the corresponding short names, in their
second columns, are self-explanatory.

The stellar mass function of the individual subclusters is taken to be canonical \citep{pk2001,pk2013} and
is drawn from a probability distribution without an upper bound.
Note that the stellar entities here are proto-stars which are yet to reach
their hydrogen-burning main sequences (a few most massive members can become MS during their infall
depending on the infall time). Furthermore, the interplay between gas accretion and dynamical processes (ejections, mergers)
during the infall and the final assembly of the main cluster continue to shape the global stellar mass function
or the initial mass function (IMF) of the final cluster \citep{kl1998}. This IMF is often observed to be
canonical for VYMCs. Note that the accretion and the dynamical processes only influence
the massive tail of the IMF and also sets the maximum stellar mass \citep{wk2004,wp2013}
of the \emph{final} cluster, as indicated in hydrodynamic calculations \citep{kl1998,giri2011}. 
The overall canonical form of the IMF as determined by the low mass stars, which contribute to most of the
stellar mass of a subcluster ($>90$\%), appears mostly due to gravitational fragmentation alone. This justifies
our adopting of the unconditional canonical IMF as a suitable representation of the proto-stellar mass function of the
subclusters, despite the exclusion of gas accretion in the calculations here. Again, the
proto-stellar mass function does not critically impact the primary results here.  

Finally, all subclusters are initially at rest w.r.t. the centre of mass (COM) of the stellar system.
While this initial condition is again an idealization, it is consistent with the results of detailed hydrodynamic
computations. In these calculations, the newly formed subclusters in the high-density filaments and at the
filament junctions typically move with velocities which are much smaller than the average turbulent velocity
of the gas cloud and hence form a sub-virial system of subclusters (see, \eg, \citealt{bate2004,bate2009}). 
As long as the initial system of clusters is sub-virial, the timescales inferred from
the calculations here would not be affected significantly.

Primordial binaries are currently excluded for the ease of computing. The presence of primordial binaries
does not affect the subcluster merging process substantially as found in test calculations (see Sec.~\ref{mrgcls}). 

\subsection{Initial gas potential}\label{initgas}

The dense residual molecular cloud, within which the proto-stars form in distinct subclusters (see Sec.~\ref{initstar}),
is represented by a background, external gravitational potential of a Plummer mass distribution. In this
way the overall dynamical effect of the molecular cloud is included. In order to compare with
the previous studies \citep{pketl2001,sb2014}, we adopt a local SFE of $\epsilon\approx1/3$.
Such SFE is also consistent with those obtained from hydrodynamic calculations including
self-regulation \citep{mnm2012,bate2013} and as well with observations of embedded systems
in the solar neighborhood \citep{lnl2003}. 

Hence, we place the geometric/density centre of the Plummer gas sphere co-incident with the COM
of the initial stellar system and take the half-mass radius of the former equal to the initial span $R_0$
of the subclusters (see Sec.~\ref{initstar}). To achieve a local (cluster-scale) SFE of $\epsilon\approx1/3$,
the total gas mass within $R_0$ is taken to be $2M_\ast$, \ie, the total matter (gas + stars) within the $R_0$-sphere is
$3M_\ast$. Since $R_0$ is equal to the half-mass radius of the Plummer gas sphere, the total
mass of the spherical gas cloud is $4M_\ast$ and, including the stars, the entire system contains $5M_\ast$ of matter. 
In principle, star clusters can also appear beyond the half-mass radius of the gas cloud although
the probability of forming them within is higher because of the higher average gas density there. Hence, the above
initial setup is again an idealization to set the intended SFE. It is also consistent with
the uniform initial distribution of the subclusters within $R_0$ since the density is nearly
uniform within the Plummer half-mass radius. Note that although the SFE is $M_\ast/3M_\ast=33$\% within $R_0$,
it is $M_\ast/5M_\ast=20$\% for entire gas cloud. These numbers are consistent with what is found in the recent
hydrodynamic calculations (see Secs.~\ref{intro} and \ref{initstar}) of star formation and are also
consistent with observations \citep{lnl2003}.

Inserting our adopted value $M_\ast=10^4\Ms$ (see Sec.~\ref{initstar}), we get a total of $3\times10^4\Ms$
(gas + stars) within $R_0$. This gives an ONC-like $\rho_g\approx6\times10^3\Ms{\rm~pc}^{-3}$
gas density for $R_0=1.06{\rm~pc}$ and $\approx1/1000$th of this for $R_0=10{\rm~pc}$ which
is appropriate for, \eg, the Taurus-Auriga complex of subclusters.

The analytic Plummer gas cloud is geometrically static but its total mass is reduced exponentially
to mimic gas expulsion as in \citet{pketl2001,sb2013}. We use this treatment
to be able to compare the present calculations with the previous ones. Dynamical
gas expulsion in SPH-based calculations have shown this treatment to yield
well comparable results \citep{gb2001}.  Note that the spherically
symmetric gas expulsion from beyond $R_0$ (which has a total mass of
$2M_\ast$) will not influence the stellar system within $R_0$ since
the former fraction of gas imparts zero gravitational force within $R_0$. Hence, the expulsion of the
entire $4M_\ast$ Plummer sphere has essentially the same dynamical effect on the stellar system as the
expulsion of just the $2M_\ast$ gas (which is appropriate for $\approx33$\% SFE) from within $R_0$.      
We address some additional issues in Sec.~\ref{discuss}.  

\section{Dynamical evolution of the subcluster systems}\label{evol}

The above initial configurations (see Table~\ref{tab:morphevol} \& \ref{tab:kfit1}) are
generated using a self-developed automated script. Its final outcome is the combined
masses, positions and velocities for all the stars in all subclusters for a particular
configuration. The script uses the {\tt MCLUSTER} code \citep{kup2011} to generate the individual Plummer
spheres.

We evolve the above initial configurations
using the state-of-the-art direct N-body integrator {\tt NBODY6} \citep{ars2003}. The initial configuration
data can be directly fed to the {\tt NBODY6} code. The spherically symmetric
background gas potential (see Sec.~\ref{initgas}) can also be initiated and varied over time
from within the N-body code. {\tt NBODY6} is a highly sophisticated N-body integrator which
uses a fourth-order Hermite scheme and particle-specific time steps to compute the trajectories
of the individual mass points. The diverging gravitational force between two close-passing masses
or that between the components of a close binary or a multiplet is handled using two- and few-body
regularization \citep{ars2003}.

{\tt NBODY6} includes the widely-used semi-analytic binary evolution scheme {\tt BSE} \citep{hur2000} to
evolve the individual stars starting from their main sequence. However, {\tt BSE} does
not include a reliable PMS evolution. Hence, in the {\tt NBODY6} calculations here, we do not activate
the stellar evolution. However, we do a few test calculations with the main sequence
evolution to qualitatively assess any effect of stellar winds. Since we are primarily interested
in the configurations at very young ages when the most massive stars are still
on their main sequence, stellar wind mass loss would not severely influence the massive stellar
spectrum and the associated dynamical consequences.

\subsection{General evolutionary properties}\label{genevol}

The generic evolution of a given configuration consists of three parts, \viz,
(a) the net infall of each subcluster towards the global minimum of the gravitational potential
well of the system, (b) the two-body relaxation of each subcluster while they
fall in and (c) the final coalescence of the subclusters to approach a system in
dynamical (virial) equilibrium. The global potential minimum of the system roughly corresponds to the
system's COM. The background Plummer gas potential (see Sec.~\ref{initgas}), if present,
is the dominant component of the global potential well. The infalling subclusters
eventually merge and form a (near) spherical core-halo stellar distribution
which is in dynamical equilibrium. During the infall and the merger
process, the potential well of the system changes with time but the final
virialized stellar distribution and the global minimum of
the final potential well must
be concentric for an energetically stable configuration. The
final virialized stellar distribution has also much smaller spatial extent than their
initial extent $R_0$ (see below).

The timescale in which
the system arrives at the smooth-profiled, single cluster configuration is
primarily determined by the initial span, $R_0$, of the subclusters.
$R_0$ determines the time taken for the subclusters to cross their orbital pericenters
(or fall through the potential minimum for strictly radial orbits). 
The background gas potential, of course, accelerates the initial infall.
Starting from rest, as in the present case (see Sec.~\ref{initstar}), the subclusters' pericenters
are typically close to the potential minimum, \ie, they populate a small
central region. This causes most subclusters to pass through
each other nearly simultaneously (or in a few lots). During this phase and
in subsequent passages, the subclusters' orbital kinetic energy (K.E.)
is dissipated into the orbital energy of the individual stars via
two-body energy exchange among the latter. This causes the subclusters' orbits to
sink towards the potential minimum (typically in a few orbits) where most of the stellar
orbits confine forming an initially irregular and substructured stellar distribution.
The redistribution of the stellar orbits of this stellar system drives
it towards its lowest energy configuration (the ``violent relaxation''
process) causing its morphology to become
increasingly spherically symmetric and the substructures to vanish. The
infall and the initial merger processes occur in the dynamical time of the
subclusters falling in the global potential well. As the newly merged
system smoothes out and approaches dynamical equilibrium, the redistribution
proceeds with the dynamical timescale of its stellar orbits.

The subclusters' infall time, $t_{in}$, nevertheless serves as a lower limit of their
merger time. An estimate of $t_{in}$ is the ``crossing time'' over $R_0$ with the typical
orbital velocity, $\sigma_{in}$, of a subcluster. An estimate of $\sigma_{in}$ is the velocity dispersion
needed to keep the system of subclusters, treated as point masses, in virial equilibrium.
The virial theorem\footnote{A system is said to be in dynamical equilibrium if its
statistical or ``macroscopic'' properties remain invariant with time or changes
quasi-statically. According to the virial theorem,
for a self-gravitating system in dynamical equilibrium having
a total K.E. $T$ and P.E. $V$, $2T=-V$.}
then gives, assuming no background gas,
$$\sigma_{in}\approx\sqrt{\frac{GM_\ast}{R_0}}.$$
Hence,
\begin{equation}
t_{in}=\frac{R_0}{\sigma_{in}}\approx\frac{R_0^{\frac{3}{2}}}{\sqrt{GM_\ast}}.
\label{eq:tin}
\end{equation}
With a background residual gas of total mass $M_g$ \emph{within $R_0$}, Eqn.~\ref{eq:tin} can be
simply generalized to
\begin{equation}
t_{in}\approx\frac{R_0^{\frac{3}{2}}}{\sqrt{GM_{tot}}}
=0.152\frac{\left(\frac{R_0}{\rm pc}\right)^{\frac{3}{2}}}
           {\left(\frac{M_{tot}}{10^4\Ms}\right)^{\frac{1}{2}}}{\rm~Myr},
\label{eq:tin2}
\end{equation}
where $M_{tot}=M_\ast+M_g$. Note that Eqn.~\ref{eq:tin2} is valid if the initial relative velocities
of the subclusters are small compared to $\sigma_{in}$ as in the present case (see Sec.~\ref{initstar}).    
 
\begin{figure*}
\includegraphics[width=12.0cm, angle=0]{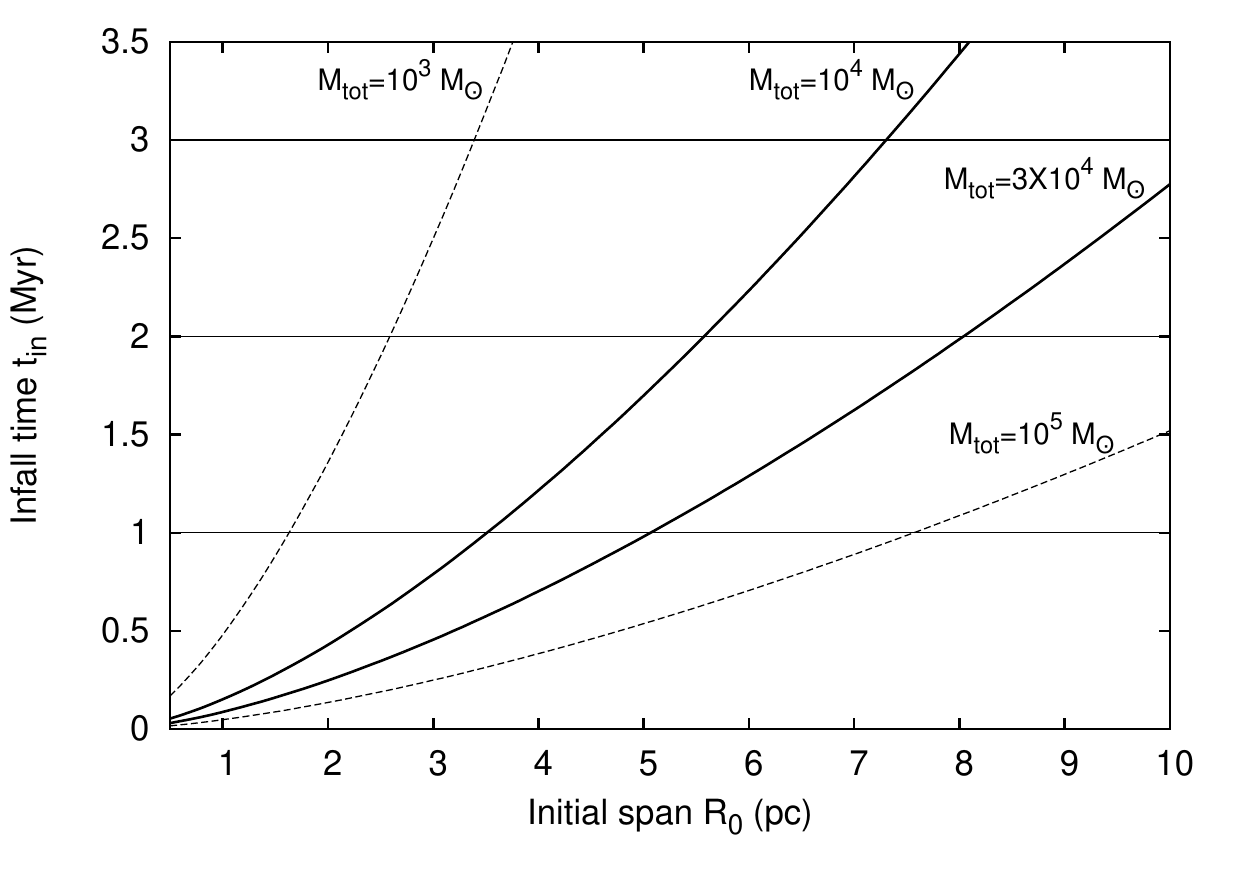}
\caption{The infall time (or the time of first arrival at orbital pericenter; see
Sec.~\ref{genevol}) of the subclusters, $t_{in}$, as a function of the radius, $R_0$,
of the spherical volume over which they are initially distributed. The curves are according
to Eqn.~\ref{eq:tin2} for different systemic mass $M_{tot}$. For the
present calculations, $M_{tot}=M_\ast=10^4\Ms$ without a residual gas and
$M_{tot}=3M_\ast=3\times10^4\Ms$ with the residual gas (see Sec.~\ref{initgas}).
These two curves are highlighted.}
\label{fig:tin}
\end{figure*}

Notably, $t_{in}$ estimates only the \emph{time of the first arrival} of the subclusters
at their pericenters. As discussed above, the time taken to merge the subclusters into
a single cluster in virial equilibrium can be substantially longer than $t_{in}$. This
additional time, in which the subclusters' orbit collapse towards the
global potential minimum and the resulting stellar system eventually arrives at a 
featureless (near) spherically symmetric configuration (see above), can be referred to
as the systemic violent relaxation time, $t_{vrx}$.
$t_{vrx}$ also scales with $R_0$ since
increasing $R_0$ increases the amount of the subclusters' orbital energy, which
needs to be dissipated (see above). This orbit-shrinkage phase of the subclusters and hence $t_{vrx}$
is too complicated to estimate analytically since the energy exchange process significantly deforms
the subclusters from spherical symmetry and hence complicates the exchange itself. A direct N-body computation,
as we do here with {\tt NBODY6} (see above), is the most reliable and accurate method
to treat the merger process after time $t\gtrsim t_{in}$.
Fig.~\ref{fig:tin} shows the
dependence of $t_{in}$ on typical masses and sizes involved in massive stellar associations. 
The thick curves correspond to the presently adopted values, \viz, $M_{tot}=(3\times)M_\ast=(3\times)10^4\Ms$
(with) without the background gas.

The final radius, $R_\ast$, of the post-merger virialized cluster can also be estimated
as follows which would be useful.
To the lowest order, the length scale of the individual subclusters can be simply related
to that of the final merged cluster in dynamical equilibrium. For purposes of estimates,
we consider all subclusters having mass $m_{cl}(0)=M_{cl}$ and virial radius $R_{cl}$. For a Plummer
cluster, $R_{cl}$ is close to its half-mass radius $r_h(0)$. The total (kinetic + potential)
energy of an individual Plummer
(sub-) cluster is,
$$E_{cl}=\frac{V}{2}\approx-\frac{M_{cl}^2}{2R_{cl}},$$ where $V$ is the total potential energy (P.E.)
of the cluster. The $G$-constant is omitted in the following derivation
since it would cancel out (see below). 

Hence, for an ensemble of $n$ clusters relatively at rest and uniformly distributed over a sphere of radius $R_0$,
as is true for the current initial conditions (see Sec.~\ref{initstar}),
the total energy of the system is
\begin{equation}
E_{ini}\approx -\frac{nM_{cl}^2}{2R_{cl}} - \frac{n^2 M_{cl}^2}{R_0}.
\label{eq:Eini}
\end{equation}
Here, $M_\ast=nM_{cl}$ is the total stellar mass. The second term in the right hand side (R.H.S.) of
Eqn.~\ref{eq:Eini} is the P.E. corresponding to $M_\ast$ distributed uniformly
over a sphere of radius $R_0$ and estimates the mutual P.E. of the subclusters. 
The corresponding K.E. term is zero since the subclusters have zero relative velocities.
If the virial radius of the final merged stellar system of mass $M_\ast$ is $R_\ast$, then the corresponding total energy is
\begin{equation}
E_{fnl}\approx-\frac{n^2M_{cl}^2}{2R_\ast}.
\label{eq:Efnl}
\end{equation}
Here we take the fraction of stars that becomes unbound during the merger process (and by the
dynamical evolution of the individual subclusters) to be negligible
as confirmed in our computations (see below). The conservation of energy implies
$$E_{ini}=E_{fnl}.$$
Then Eqns.~\ref{eq:Eini} \& \ref{eq:Efnl} give, after dividing by $n^2M_{cl}^2$,
\begin{equation}
\frac{1}{2R_\ast}\approx\frac{1}{2nR_{cl}}+\frac{1}{R_0},
\label{eq:Rmrg}
\end{equation}
which is independent of the systemic stellar mass $M_\ast$. With a time-independent
external potential, Eqn.~\ref{eq:Rmrg} also holds true. Note that in the above calculation
we ignore the corrections in the total energy due to the mutual tidal forces
and hence Eqn.~\ref{eq:Rmrg} is correct up to only
the leading orders.

Table~\ref{tab:morphevol} summarizes the evolution (in Myr) of the subcluster systems 
of type A and B (see Sec.~\ref{initstar}) initiating with increasing $R_0$ and
in presence (for $\epsilon\approx0.3$; see Sec.~\ref{initgas}) and absence of a gas potential.
These are computed using {\tt NBODY6} (see above).
For the ease of description, the evolving morphologies of the stellar systems are divided
into four categories as given in Table~\ref{tab:morphdef}. As seen in
Table~\ref{tab:morphevol}, all computed configurations initiate as SUB
and evolve via the intermediate CHas phase to the final CH cluster in dynamical
equilibrium. In the CHas state, the stellar system has an overall core-halo structure
but it is out of equilibrium and approaches dynamical equilibrium on the timescale
of its typical stellar orbits\footnote{This timescale is of the order of the crossing time of
the final virialized cluster.}.

\begin{figure*}
\centering
\includegraphics[width=16.8cm, angle=0]{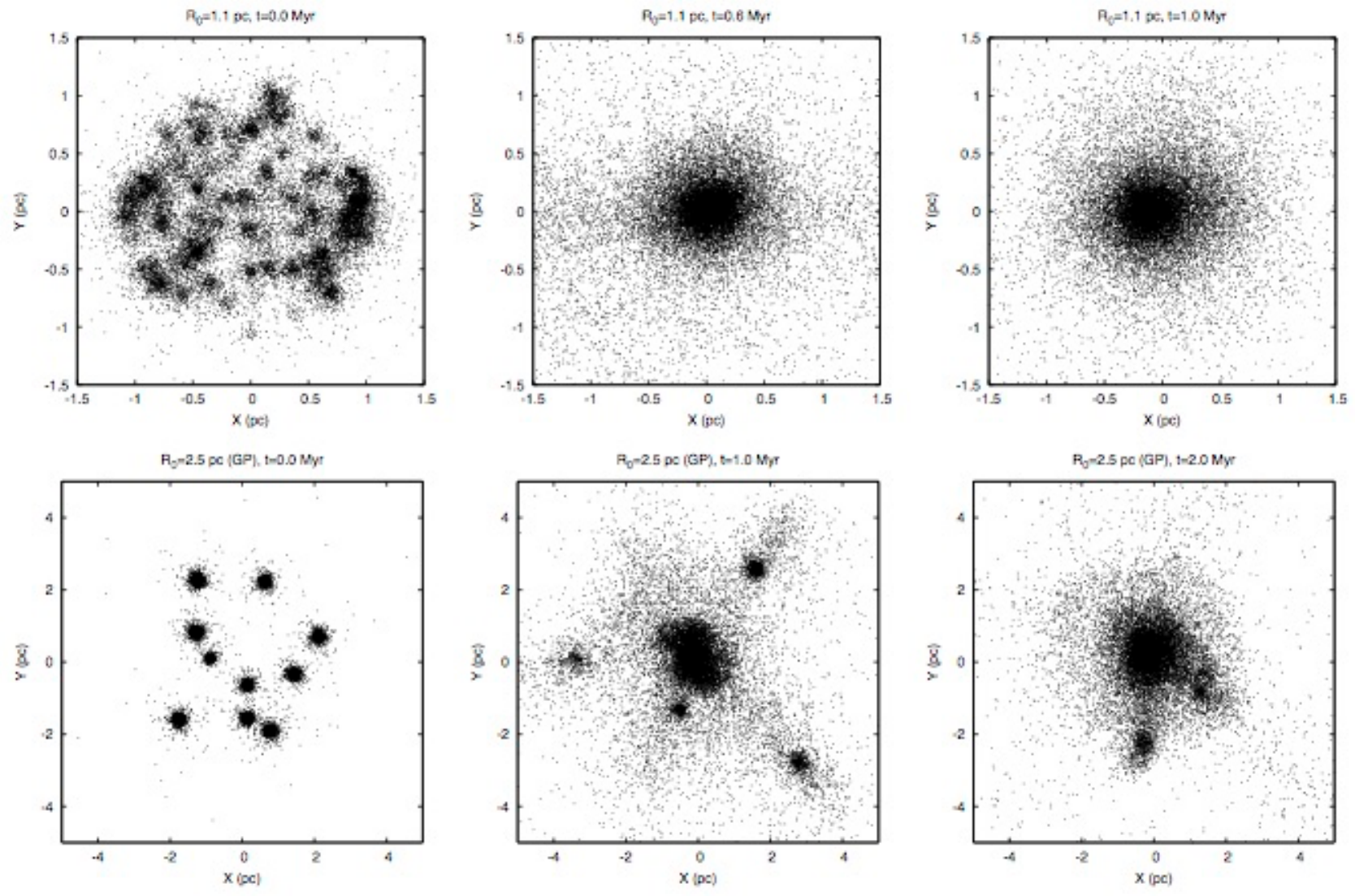}
\caption{Examples of evolution of ``compact'' configurations, \ie, configurations computed here which
are initially distributed over spherical volumes of radii $R_0\leq2.5$ pc. These constitute all
the initial systems abbreviated as $\ast$I$\ast$ and $\ast$II$\ast$ in
Tables~\ref{tab:morphevol} and \ref{tab:kfit1}. Here, the evolutions of
the systems B-Ia and (upper 3 panels) and A-IIa (lower 3 panels) are shown.
``GP'' indicates the presence of a gas potential (Sec.~\ref{initgas}).}
\label{fig:snaps_egevol}
\end{figure*}

As expected, configurations with smaller $R_0$
tend to merge quicker, as can be read from Table~\ref{tab:morphevol}. For $R_0\lesssim 1{\rm~pc}$,
the system attains a CH structure in $t\lesssim1{\rm~Myr}$ without a gas potential. This is
demonstrated in Fig.~\ref{fig:snaps_egevol} (panels 1-3) where the initial system falls in from $R_0=1.1{\rm~pc}$.
On the other hand, initially wider configurations in Table~\ref{tab:morphevol} are all still SUB
at $t=1{\rm~Myr}$ even with the background gas potential and most of them do not attain
the CH phase even in 2 Myr. An example is shown in Fig.~\ref{fig:snaps_egevol} (panels 4-6).

\begin{figure*}
\vspace{-0.3 cm}
\centering
\includegraphics[width=15.0cm, angle=0]{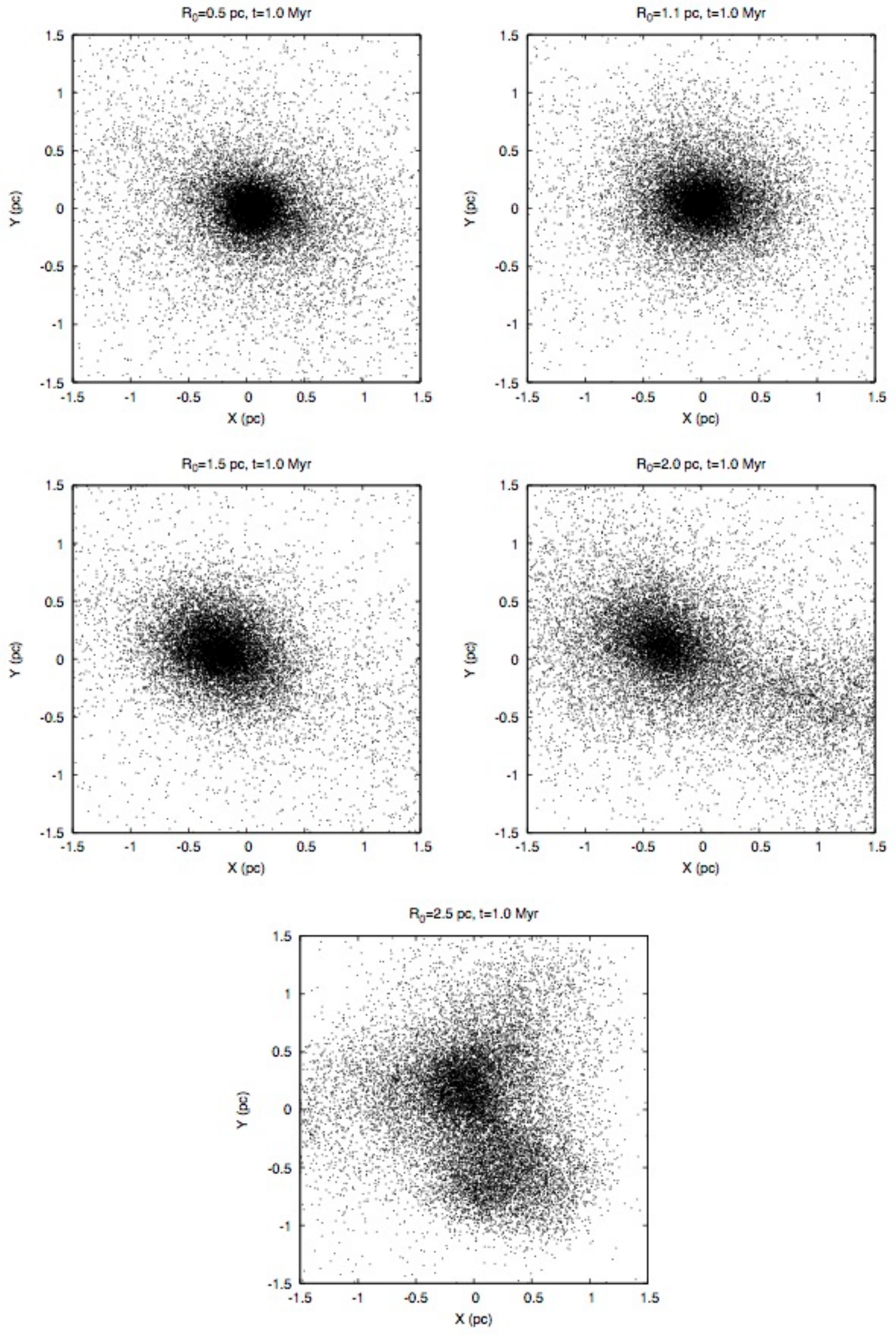}
\caption{Configurations obtained at $t\approx1$ Myr with increasing initial span $R_0$ for gasless
``compact'' Type-A systems with $r_h(0)=0.3$ pc, namely, the systems A-0.5Ib, A-Ib, A-1.5Ib,
A-2.0Ib and A-IIb (panels 1-5 respectively, numbered left-to-right, top-to-bottom;
see Tables~\ref{tab:morphevol} and \ref{tab:kfit1}).
With increasing $R_0$, the system's morphology at $t\approx1$ Myr changes from being
near-spherical core-halo (CH; panels 1,2), asymmetric core-halo (CHas; panels 3,4) to substructured
(SUB; panel 5). For $R_0\gtrsim2$ pc (panels 4,5), the stellar system is still well in the process of
merging at $t\approx1$ Myr after the subclusters' first
pericenter crossings (\ie, it is in the violent relaxation phase
$t_{in}<t<t_{in}+t_{vrx}$; see Sec.~\ref{genevol}). 
}
\label{fig:snaps83_1myr}
\end{figure*}

\begin{figure*}
\centering
\includegraphics[width=17.1cm, angle=0]{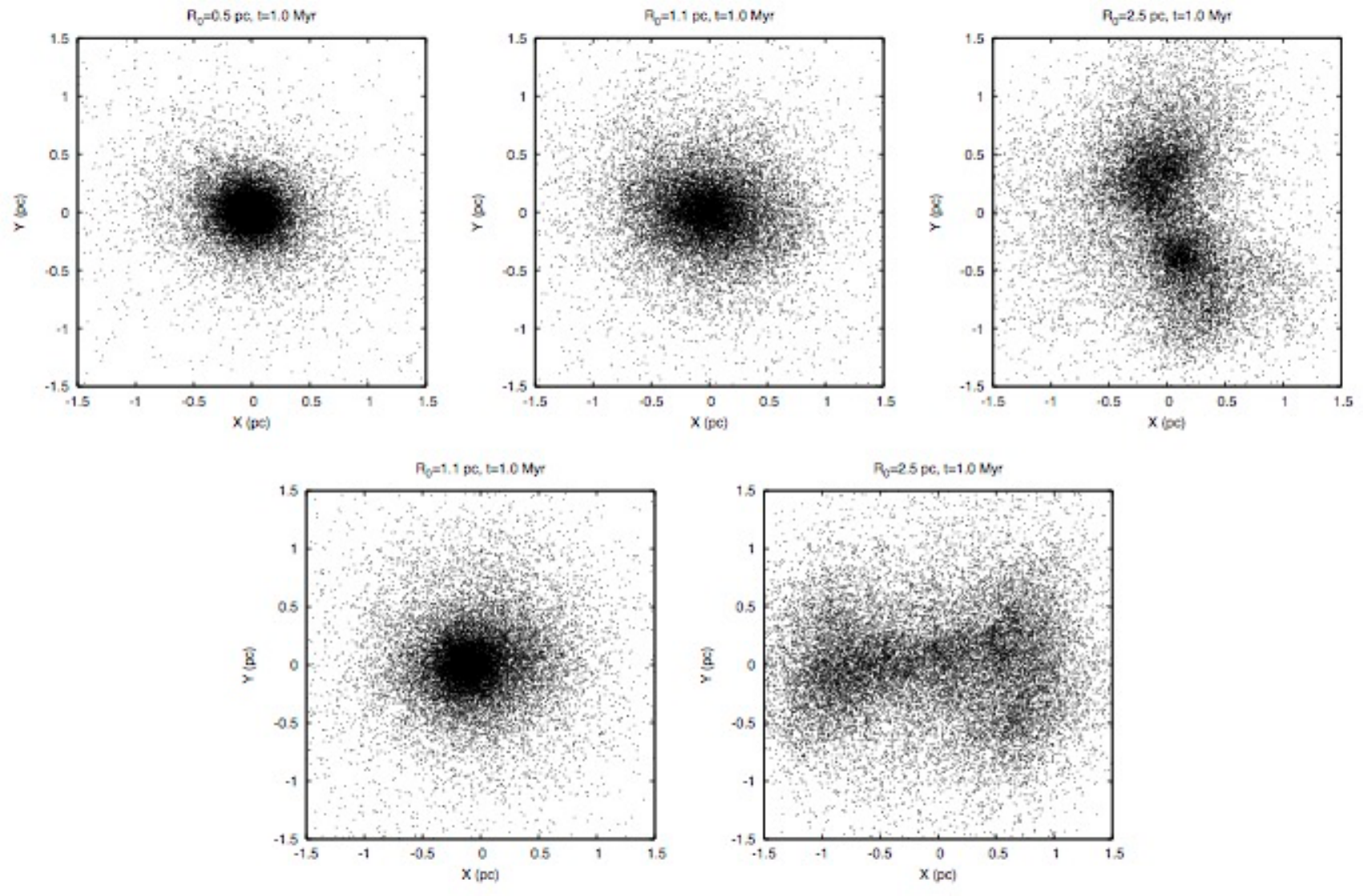}
\caption{
Configurations at $t\approx1$ Myr with increasing initial span $R_0$ for
gasless ``compact'' Type-A (upper three panels)
and Type-B (lower two panels) systems with
$r_h(0)=0.1$ pc. These are the systems A-0.5Ia, A-Ia, A-IIa, B-Ia and B-IIa
respectively (see Tables~\ref{tab:morphevol} and \ref{tab:kfit1}).
As in Fig.~\ref{fig:snaps83_1myr}, the configurations change through CH, CHas and SUB
with increasing $R_0$. For $R_0=2.5$ pc, the stellar system is still in the
merging process at $t\approx1$ Myr, likewise in Fig.~\ref{fig:snaps83_1myr}.  
}
\label{fig:snaps8n9_1myr}
\end{figure*}

Fig.~\ref{fig:snaps83_1myr} shows the snapshots at $t\approx1{\rm~Myr}$ for a set of A-type configurations 
(systems A$\ast$b; see Tables~\ref{tab:morphevol} \& \ref{tab:kfit1}) falling from increasing
$R_0$ (without background gas potentials). With increasing
$R_0$, the morphology at $1{\rm~Myr}$ changes from being CH, CHas to SUB. For $R_0\gtrsim2{\rm~pc}$,
the stellar system at $1{\rm~Myr}$ substantially deviates from spherical symmetry (and dynamical
equilibrium). Fig.~\ref{fig:snaps8n9_1myr} demonstrates the same for A- and B-type configurations with the
more compact initial subclusters (systems A$\ast$a and B$\ast$a; see Tables~\ref{tab:morphevol} \& \ref{tab:kfit1}).
Note that the infall time, $t_{in}$, for $R_0\approx2{\rm~pc}$ ($M_{tot}\approx10^4\Ms$) is much less than
1 Myr, as can be read from Fig.~\ref{fig:tin}. This demonstrates the significance of the
systemic violent relaxation time, $t_{vrx}$, beyond $t_{in}$ (see above) which delays the appearance of
the virialized merged cluster for $t>1$ Myr, in this case.  

Interestingly, the presence of a gas potential (Sec.~\ref{initgas}) does not necessarily facilitate the approach
to the CH morphology, as can be seen from Table~\ref{tab:morphevol}. Although the gas potential
causes the subclusters to make the first passage through each other (near their orbital
pericenters) quicker (\ie, makes $t_{in}$ smaller; see  Eqn.~\ref{eq:tin2}),
they also approach faster than what they would have under just their
mutual gravity.
Hence a larger number of orbits are needed to dissipate the subclusters' orbital K.E. to that
of the stars (see above). In other words, the addition of a gas potential tends to lengthen $t_{vrx}$
without increasing $R_0$. 
For $R_0\gtrsim2{\rm~pc}$, it takes longer to arrive at the CH state
with the gas background than without as seen in the present N-body calculations. Hence,
the ``conveyer belt'' mechanism \citep{longm2014}, in fact, tends to delay
the assembly of the final star cluster.
This, of course, depends on the gas mass (potential) in the system relative to
the stellar mass. A detailed quantification of the role of the background gas in the subcluster
merger timescale is reserved for a future paper.

Furthermore, systems with similarly massive but less compact subclusters generally take longer to merge as can be
inferred from Table~\ref{tab:morphevol}. The times at which the subclusters with $r_h(0)=0.1{\rm~pc}$
combine to form an equilibrated CH structure, those with $r_h(0)=0.3{\rm~pc}$ are still in CHas state.
More compact subclusters have higher stellar densities which cause their orbital energy dissipation
to be more efficient as they pass through each other at the beginning of the merger process (see above).  

\begin{figure*}
\centering
\includegraphics[width=15.0cm, angle=0]{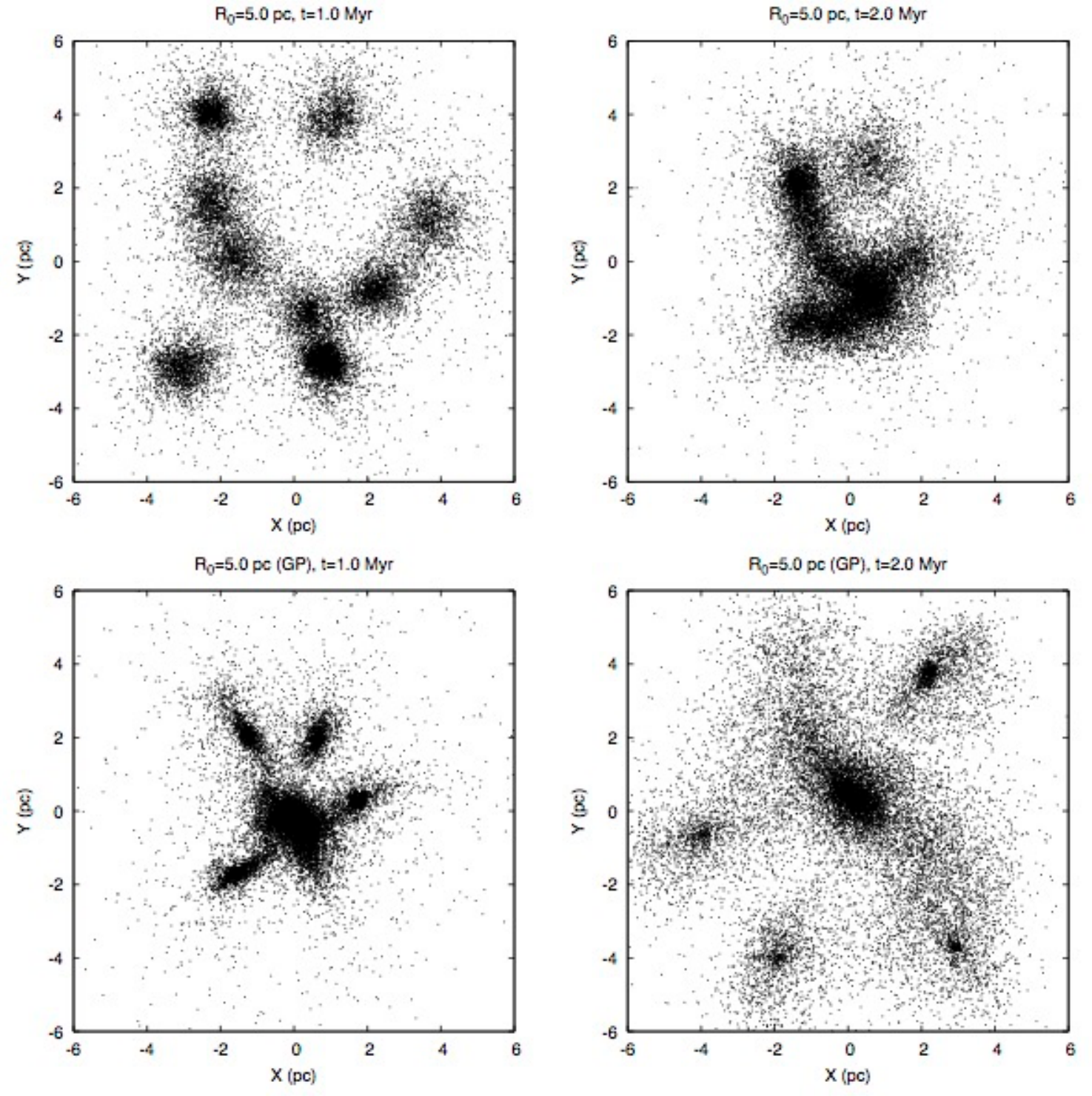}
\caption{
Evolution for the ``extended'' system A-IIId ($R_0=5$ pc; see Table~\ref{tab:morphevol}).
As expected, the infall of the subclusters proceeds much more slowly than that for the compact systems
(\cf, Figs.~\ref{fig:snaps_egevol}). The system is still highly substructured (SUB) at
$t\approx2.0$ Myr both in absence (panels 1,2, numbered left-to-right, top-to-bottom)
and presence (panels 3,4) of a gas potential.
In presence of the gas potential, the subclusters are close to 
the first arrival at their pericenters (\ie, $t\approx t_{in}$;
see Sec.~\ref{genevol}) at $t\approx1$ Myr (panel 3) while
this takes $t\approx2$ Myr without the gas (panel 2). This is consistent with Fig.~\ref{fig:tin}.
For the evolution with gas potential, the subclusters make their
first passage through each other during $1-2$ Myr which is why their configuration at 2 Myr
(panel 4) is more extended than that in 1 Myr (panel 3). Note that the gas potential
tidally elongates the subclusters significantly (also see Fig.~\ref{fig:snaps7_R10}).
}
\label{fig:snaps7_R5}
\end{figure*}

\begin{figure*}
\centering
\includegraphics[width=15.0cm, angle=0]{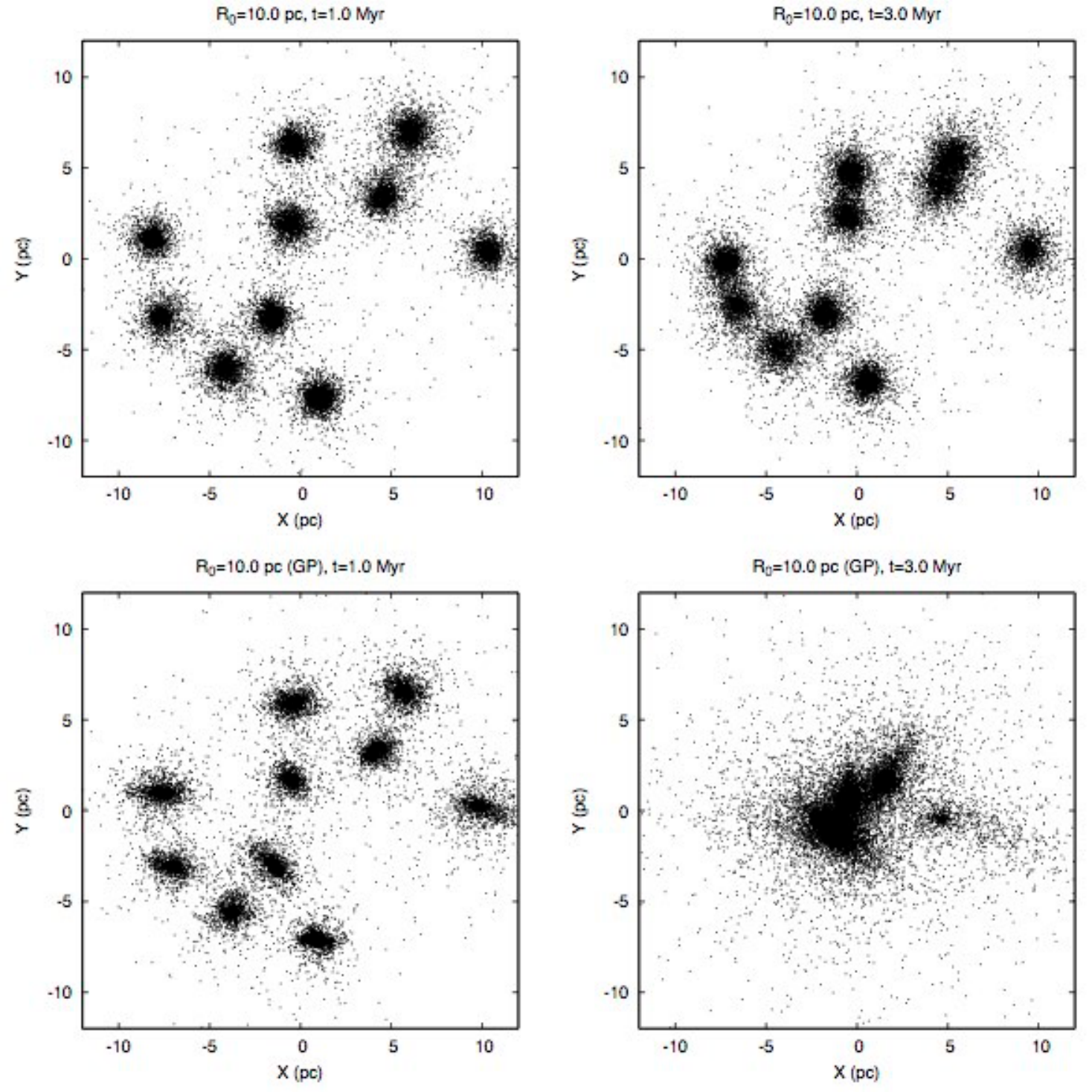}
\caption{
Evolution for the ``extended'' system A-IVd ($R_0=10.0$ pc; see Table~\ref{tab:morphevol})
without (panels 1-2, numbered left-to-right, top-to-bottom)
and with (panels 3-4) gas potential as in Fig.~\ref{fig:snaps7_R5}.
In both cases, the system still remains highly substructured at $t\approx3$ Myr. With the
gas potential, the subclusters arrive at their pericenters for the first time at
$t\approx3$ Myr (panel 4) but are far from reaching there at that time without the gas (panel 2),
being consistent with Fig.~\ref{fig:tin}.
}
\label{fig:snaps7_R10}
\end{figure*}

Figs.~\ref{fig:snaps7_R5} \& \ref{fig:snaps7_R10} show examples of infall from $R_0=5{\rm~pc}$
and $10{\rm~pc}$ respectively comparing between the cases with and without the background gas.   
The systems are still highly SUB at 2 \& 3 Myr respectively (the $R_0=5{\rm~pc}$ systems become
CH by 3 Myr). The faster initial collapse in presence of the gas potential can also be
vividly seen in these figures. In panel 3 of Fig.~\ref{fig:snaps7_R5} ($R_0=5{\rm~pc}$), \ie, with the
gas potential, the stellar system is close to its first mutual (pericenter) passage at $t\approx t_{in}\approx1$ Myr
and spreads out further after a Myr (panel 4). The first mutual passage occurs at
$t\approx t_{in}\approx2$ Myr without the gas potential for the same stellar system (panel 2).
These times are close to what one expects from the analytic estimate of $t_{in}$ for
$R_0\approx5{\rm~pc}$ in Fig.~\ref{fig:tin} (\cf the highlighted curves in this figure).   
Similarly, for $R_0=10{\rm~pc}$, the stellar system is close to its first pericenter at $t\approx3$ Myr
(panel 4 of Fig.~\ref{fig:snaps7_R10}) with the gas potential but far from reaching there
without it (panel 2), as can be expected from Fig.~\ref{fig:tin}.    
Notably, the tidal field due to the background potential elongates the subclusters as can be
seen in the panels 3 \& 4 of Figs.~\ref{fig:snaps7_R5} \& \ref{fig:snaps7_R10}.

In all the calculations reported here (both without and with gas potential), a negligible fraction of stars escape the system
during the infall and the merger process. In other words, the total bound stellar mass $M_\ast$
remains nearly unaltered as the system evolves from its SUB to its CH morphology. 
In the panels of Figs.~\ref{fig:snaps_0myr} - \ref{fig:snaps7_R10} only those stars are plotted
which are gravitationally bound w.r.t. the entire system. 

\subsection{Structure of the assembled star cluster: comparison with NGC 3603 young cluster}\label{mrgcls}

To obtain an HD97950-like star cluster by hierarchical merging of subclusters, the necessary
but not sufficient condition is to arrive at a CH configuration in $t\lesssim1{\rm~Myr}$.
The calculations described in Sec.~\ref{genevol} imply that
to have a CH morphology at 1 Myr, one should have $R_0\lesssim2{\rm~pc}$ 
with or without a gas potential (see Table~\ref{tab:morphevol} and Fig.~\ref{fig:snaps83_1myr}).
As discussed there, the presence of the gas potential would delay the appearance of the
CH configuration. How does this final cluster compare with the observed HD97950 cluster?

Table~\ref{tab:harafit} lists the best-fit parameters of the King surface mass-density ($\Sigma_M$) profile
\citep{king1962,hh2003} fitted to the observed \citet{hara2008} profile of HD97950.
The set of values of the King concentration
parameter $k$ and the core-radius $r_c$, as in the first line of Table~\ref{tab:harafit},
is also quoted by \citet{hara2008}\footnote{$1''\approx0.03$ pc at the distance of NGC 3603.}. 
However, for the two innermost annuli of this profile, $\Sigma_M$
fluctuates considerably (see fig.~14 of \citealt{hara2008}) indicating possible
systematic and/or crowding effects. Hence we also quote the best-fitted King parameters
in Table~\ref{tab:harafit} (with similar $\chi^2$) excluding either of the two innermost points, giving the
limits of $k$. Note that the King tidal cutoff radius, $r_t$, is consistent with being very large for
all the three fits and is nearly uncorrelated with $k$ and $r_c$, implying that HD97950
is untruncated. In other words the cluster under-fills
its Roche lobe (Jacobi radius) in the Galactic potential.

\begin{figure*}
\vspace{-0.3 cm}
\centering
\includegraphics[width=16.0cm, angle=0]{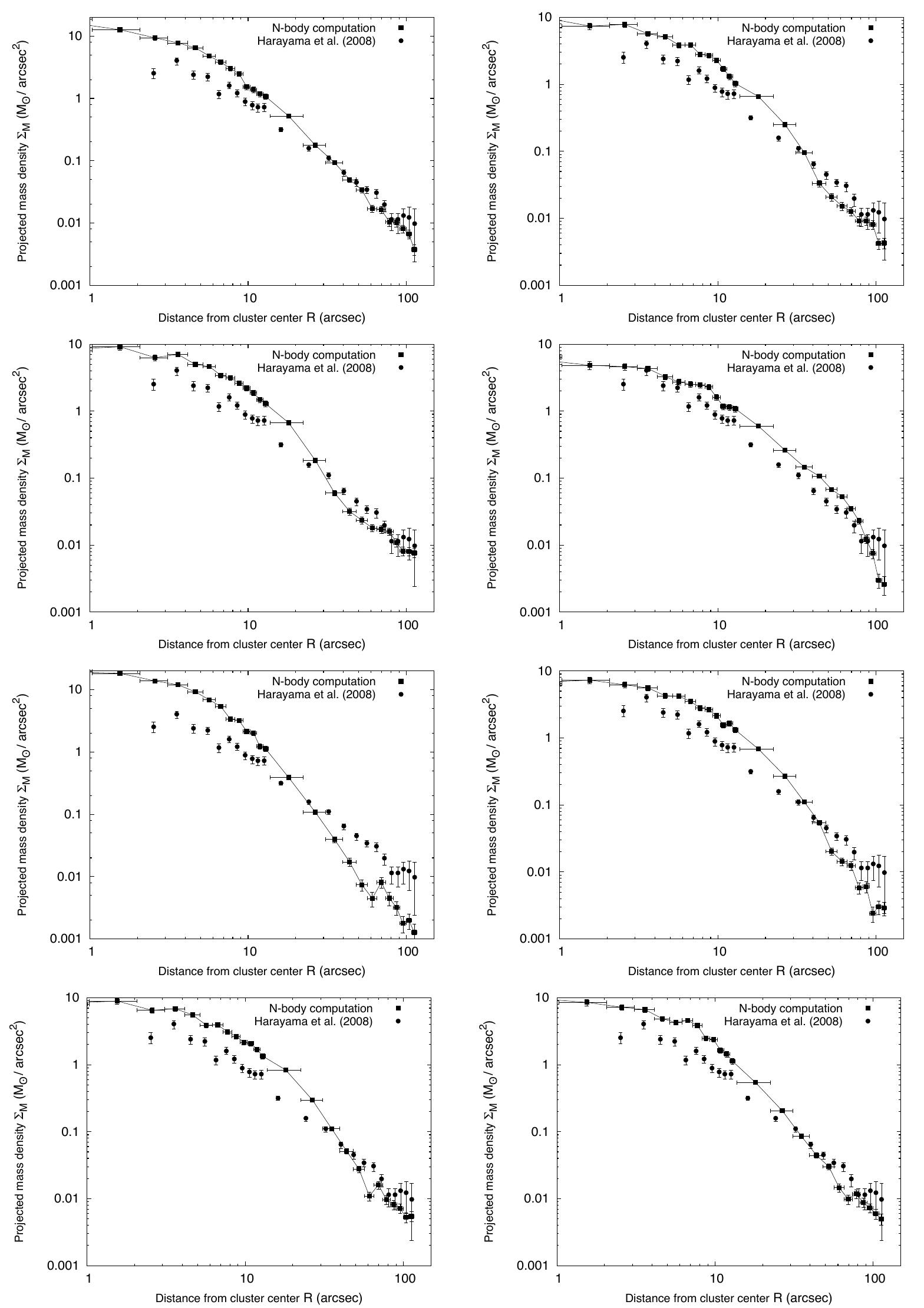}
\vspace{-0.3 cm}
\caption{
Surface mass-density profiles at $t\approx1$ Myr for those
computed configurations (filled squares connected with solid line)
which evolve to form a star cluster with near-spherical core-halo structure
(the CH-type morphology; see Table~\ref{tab:morphdef}) within $t<1$ Myr,
in absence of a background gas potential. These systems are A-0.5Ib, A-Ib, A-1.5Ib,
A-2.0Ib, A-0.5Ia, A-Ia,
B-Ic and B-Ia (Table~\ref{tab:kfit1}) as shown in panels 1-8 respectively
(numbered left-to-right, top-to-bottom). All these computed profiles are significantly more
compact and centrally overdense
than the observed one for HD97950 (\citealt{hara2008};
filled circles). This is also evident from Tables~\ref{tab:harafit}
and \ref{tab:kfit1}. For HD97950 $1''\approx0.03$ pc.
}
\label{fig:dryprofs_1myr}
\end{figure*}

Table~\ref{tab:kfit1} shows the King best-fit parameters at $t\approx1$ Myr for those gas-free systems
which arrive at a CH morphology in $t\lesssim1$ Myr, \ie, for systems with $R_0\lesssim2$ pc.
Although (near) spherical star clusters are formed by 1 Myr in these cases, they have much
higher central concentrations ($k$) and/or larger cores ($r_c$) than what is observed for HD97950
(\cf Table~\ref{tab:harafit}). This is demonstrated in Fig.~\ref{fig:dryprofs_1myr}. At $t\approx1{\rm~Myr}$,
the overall density profiles of these merged CH clusters follow neither a Plummer nor a King profile strictly.
Due to more freedom in parameters, the King surface density profile still fits reasonably with these configurations
except over their outermost parts (see Table~\ref{tab:kfit1}). Such fitted King profiles are usually truncated
(\ie, have finite $r_t$, unlike what is observed; see above)
and fall below the computed profiles in their outer
zones, typically beyond the vivid kink in their profiles (see Fig.~\ref{fig:dryprofs_1myr}) at around 40 - 60 arcsec
(1.2 pc - 1.8 pc). This indicates that these outermost ``halo'' regions of these newly assembled systems
are still in the process of integrating with their main clusters. Furthermore, in the outer radii,
the computed density profiles typically fall below the observed profile as seen in
Fig.~\ref{fig:dryprofs_1myr}, implying that the former are more compact. Note that in Fig.~\ref{fig:dryprofs_1myr},
all the computed and the observed profiles use the same radial bins and stellar mass range for direct comparison.

Note that the computed clusters in Table~\ref{tab:kfit1} are for
$M_\ast\approx10000\Ms$ which is the lower mass limit of HD97950. As shown in Sec.~\ref{genevol} (see Eqn.~\ref{eq:Rmrg}),
the length scale of the final merged cluster, $R_\ast$, is nearly independent of the total stellar mass $M_\ast$.
Hence, a stellar system with larger $M_\ast$ would combine to a cluster of the
same size, \ie, it would be even more overdense compared to the observed HD97950
\footnote{A system with larger $M_\ast$ would also take longer to merge since a
larger amount of orbital energy needs to be dissipated (see Sec.~\ref{genevol}).}. The appearance of the
centrally overdense assembly is irrespective of the initial mode of subclustering as
it occurs for both type A and B systems with a range of masses and sizes of the subclusters (see
Table~\ref{tab:kfit1}).

As such, according to  
Eqn.~\ref{eq:Rmrg}, more extended (larger $R_\ast$) and hence less centrally dense assembly can be obtained
starting with more extended subclusters (larger $R_{cl}$), for a given $R_0$. However, our
choice of the highly compact initial subclusters is dictated by the observed thinness of the
dense molecular gas filaments (see Sec.~\ref{initstar}). Also, under typical conditions,
one would have $R_0>>R_{cl}$, making the first term in the R.H.S. of Eqn.~\ref{eq:Rmrg}
to be dominant (unless $n$ is very large). Hence, if one relaxes the $t<1$ Myr condition, \ie,
considers larger $R_0$, one would still obtain similarly dense assemblies. Finally, according
to Eqn.~\ref{eq:Rmrg}, $R_\ast$ is independent of the background gas potential if the potential
remains static during the infall and the subsequent merger process.

\begin{figure*}
\centering
\includegraphics[width=16.0cm, angle=0]{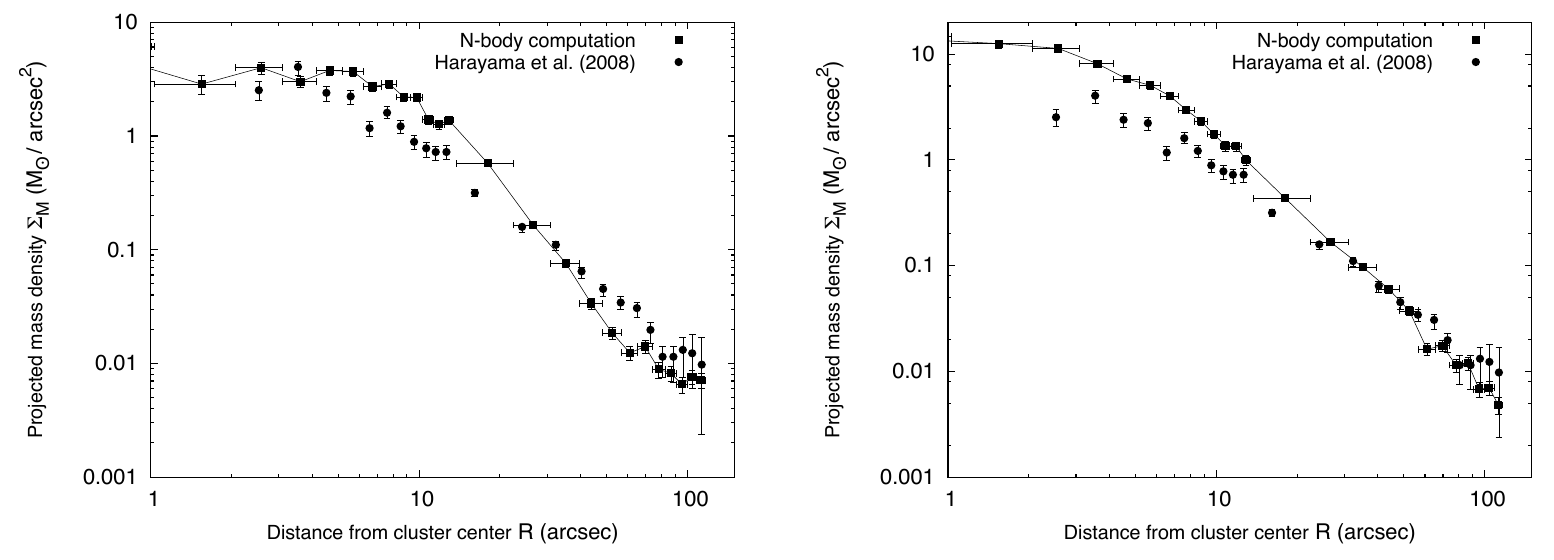}
\caption{
Similar as Fig.~\ref{fig:dryprofs_1myr} but for a system containing primordial binaries (left, system A-Ib-pb)
and a system including stellar evolution (right, system A-0.5Ib-se). Here, all legends are the same
as in Fig.~\ref{fig:dryprofs_1myr}. Neither the dynamical heating effect
due to primordial binaries nor the wind mass loss from the stars can prevent the merged structure from being
substantially overdense and more compact compared to the observed one. This is also evident from Tables~\ref{tab:harafit}
and \ref{tab:kfit1}.
}
\label{fig:dryprofs2_1myr}
\end{figure*}

The newly assembled cluster can be more extended by including primordial binaries which
``heats'' the cluster via super-elastic binary-single star and binary-binary encounters \citep{hh2003}.
The merged cluster can also be heated and expanded by mass loss due to the stellar winds.  
As shown in the last section of Table~\ref{tab:kfit1}, the stellar wind has a negligible effect
on the merged CH system and the primordial binaries expand the cluster but insufficiently;
\cf values in Table~\ref{tab:kfit1} for similar configurations but without stellar evolution and primordial binaries.
This is demonstrated in Fig.~\ref{fig:dryprofs2_1myr}.
As mentioned earlier, {\tt NBODY6} does not include PMS evolution, so we use its default
MS wind which is strongest for the massive stars
(the most massive star in the system is of $\approx80\Ms$). As for the primordial binaries,
we use the ``birth period distribution'' \citep{pk1995b,mk2014} as in \citet{sb2014}, for each initial
subcluster. Such initial binary population contains $\approx40$\% hard binaries.

\subsubsection{Calculations including gas expulsion}\label{mrgexp}

One way to dramatically expand a star cluster, however, is to subject it to a substantial gas expulsion
on a timescale of the order of its dynamical time, as discussed in Sec.~\ref{intro}. We repeat a subset of
the calculations in Table~\ref{tab:kfit1} by initiating them
in presence of Plummer gas potentials as described in Sec.~\ref{initgas}.
The gas potential is then diluted exponentially with time, mimicking gas expulsion (see Sec.~\ref{initgas}). The
timescales defining the potential depletion are taken to be the same representative values as in \citet{sb2014}. 
Specifically, the gas cloud starts to expel after a delay time $\tau_d\approx0.6{\rm~Myr}$,
after it is ionized by the UV radiation from the massive stars, with the HII sound
speed of $v_g\approx10{\rm~km~s}^{-1}$. This gives a decay timescale of $\tau_g=R_0/v_g$. As discussed in
\citet{sb2014}, these timescales are only representatives but consistent with recent theoretical
studies on ultra-compact HII regions. As discussed in Sec.~\ref{initgas}, the total gas mass
within $R_0$ is chosen to be $2M_\ast$ so that the local SFE $\approx33$\%. This SFE is consistent with both
theoretical studies and observations.

\begin{figure*}
\centering
\includegraphics[width=16.0cm, angle=0]{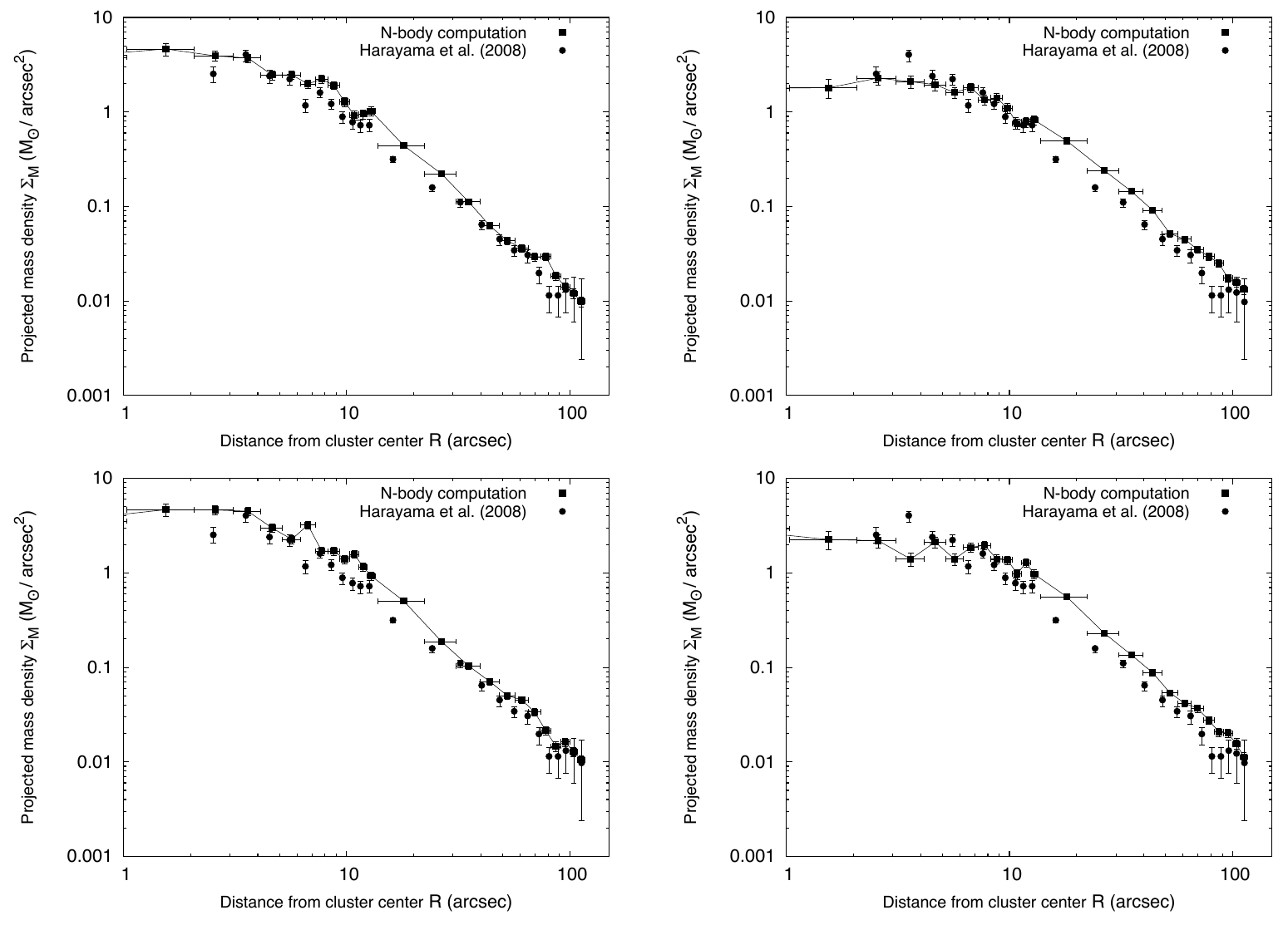}
\caption{
Surface mass-density profiles at $t\approx1$ Myr for computed post-gas-expulsion configurations.
Here, the systems evolve in a background residual gas potential (see Secs.~\ref{initgas} \& \ref{mrgexp})
to form a star cluster with near-spherical core-halo structure
(the CH-type morphology, see Table 1) within $t<1$ Myr followed by residual gas dispersal at
$\tau_d\approx0.6$ Myr. The legends are the same
as in Fig.~\ref{fig:dryprofs_1myr}. Panels 1-4 (numbered left-to-right, top-to-bottom)
are for the configurations A-Ib, A-Ia, B-Ia and A-Ib-pb
respectively (Table~\ref{tab:kfit2}).
All these profiles agree well with the observed one for HD97950 as is
also evident from Tables~\ref{tab:harafit} and \ref{tab:kfit2}.
}
\label{fig:GPprofs_1myr}
\end{figure*}

The columns 1 \& 2 of Table~\ref{tab:kfit2} give the 4 computed configurations which are subjected to
residual gas dispersal in the above way. The final calculation
also includes primordial binaries (see above). All these configurations arrive at a CH state by
$t\lesssim0.4$ Myr, \ie, well before the commencement of gas dispersal at $t\approx0.6$ Myr
\footnote{Test calculations are also done when the gas expulsion
is initiated as early as during the initial infall process, well before the CH phase.
This typically leads to the dispersal of most of the subclusters instead of merging
into a single cluster.}.
Fig.~\ref{fig:GPprofs_1myr} shows the corresponding
stellar mass-density profiles at $t\approx1$ Myr. As demonstrated there, they agree reasonably
with the observed profile of HD97950 \citep{hara2008}, particularly in the inner regions. The King best-fit parameters
for these profiles are given in Table~\ref{tab:kfit2} which agree fairly with those of
the observed ones. In particular, these profiles are consistent with being untruncated monolithic
profiles (\ie, they have large $r_t$), as observed, and unlike those obtained in the absence of
gas dispersal (\ie, without any gas potential; see above). See the notes in Table~\ref{tab:kfit2}
for additional clarifications. Note that in Fig.~\ref{fig:GPprofs_1myr}, the ``natural'' matchings
with the observed profile are obtained \emph{by simply overlaying it with the computed profiles at 1 Myr
without any scaling or fitting}, as in \citet{sb2014}. As in  Figs.~\ref{fig:dryprofs_1myr}
and \ref{fig:dryprofs2_1myr}, the same radial bins and stellar mass range are used to construct
the observed and the computed profiles in Fig.~\ref{fig:GPprofs_1myr}.    

\begin{figure*}
\centering
\includegraphics[width=16.0cm, angle=0]{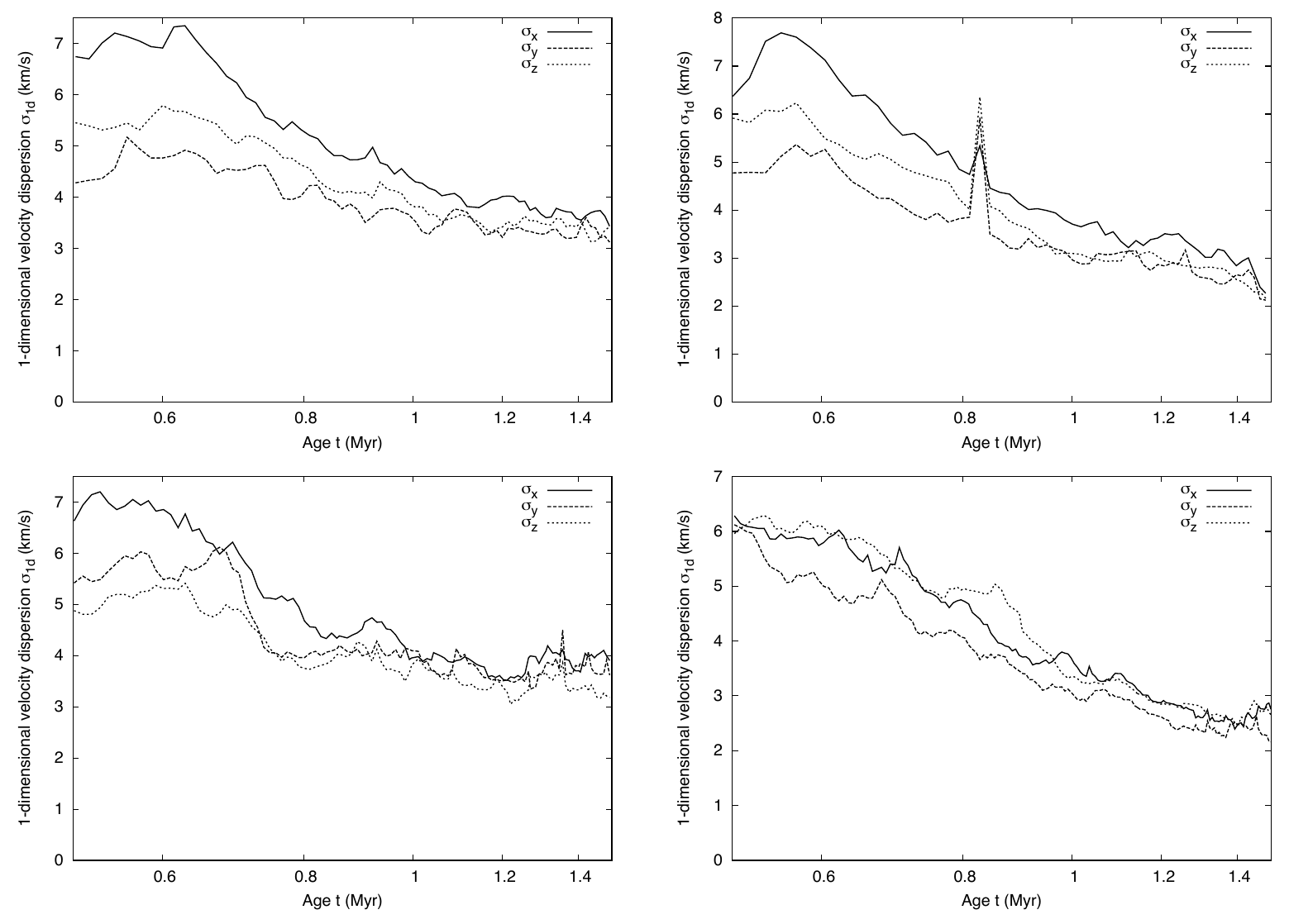}
\caption{Evolution of 1-dimensional velocity dispersion $\sigma_{\rm 1d}$ for the calculations
corresponding to Fig.~\ref{fig:GPprofs_1myr}. The values of $\sigma_{\rm 1d}$ are obtained
over the stellar mass range $1.7\Ms-9.0\Ms$ for the central $R\leq0.5$ pc of each cluster
as in \citet{roch2010}. 
}
\label{fig:vprofs_GP}
\end{figure*}

Fig.~\ref{fig:vprofs_GP} shows the evolution of 1-dimensional velocity dispersion, $\sigma_{\rm~1d}$,
for the computations corresponding to Fig.~\ref{fig:GPprofs_1myr}. In all cases, the $\sigma_{\rm~1d}$ components within
the central $R<0.5$ pc ($R\lesssim15''$) of the star cluster,
for $1.7-9.0\Ms$ stars, lie between $3.0 - 4.0{\rm~km~s}^{-1}$ at $t\approx1{\rm~Myr}$. These values are somewhat smaller than what
is observed for the same region and stellar mass range in HD97950 \citep{roch2010}, \viz,
$\sigma_{\rm~1d}=4.5\pm0.8{\rm~km~s}^{-1}$ (mean over two orthogonal components). \citet{pang2013} report
a substantial spread among the $\sigma_{\rm~1d}$ components, \viz, $4.5-7.0{\rm~km~s}^{-1}$ which
is larger than what is obtained in the above calculations. This is unlike the case of monolithic
cluster-formation model in \citet{sb2014}
where the $\Sigma_M$ profile as well as the $\sigma_{\rm~1d}$s from the same computed model agree
with those observed for HD97950.

In fact, the central $\sigma_{\rm~1d}$s in the present
calculations are smaller than those in \citet{sb2014} throughout the evolution. This can be
attributed to the smaller central concentrations of the assembled pre-gas-expulsion CH clusters here,
than that of the \citet{sb2014} monolithic embedded cluster (which corresponds to $k\approx40$, $r_c\approx2''$ ).
Note that the total stellar mass of the pre-gas-expulsion cluster $M_\ast\approx10000\Ms$
for both cases. To obtain a similarly concentrated assembled cluster, the initial subclusters
should have been much more concentrated (Eqn.~\ref{eq:Rmrg}). The smallest sized subclusters
in the calculations in Table~\ref{tab:kfit2} have scale length $r_h(0)\approx0.1{\rm~pc}$;
this is the peak of the narrow distribution of the widths of molecular gas filaments \citep{andr2011}.
Hence taking all subclusters substantially more compact would not correspond to a realistic situation.

For the sake of argument, even if one considers such an initial condition, the subsequent gas expulsion
would yield a cluster which is centrally overdense w.r.t. HD97950 even for its above lower mass limit.  
This is because the currently computed models in Table~\ref{tab:kfit2} already yield the
right central density and the latter quantity scales with its pre-gas-expulsion value. 
As argued in \citet{sb2014}, moderate variations of the SFE or the gas dispersal timescales would
not mend this since at $t\approx1$ Myr all stars are still in the
central region of the expanding system and are always intercepted in the density profile.
A similar line of argument applies if one considers a larger $M_\ast$.
Notably, the expansion of the merged cluster after the gas expulsion and hence the corresponding
decline of $\sigma_{\rm~1d}$ occur to a smaller extent in the present calculations than those in \citet{sb2014}
(\cf Fig.~\ref{fig:vprofs_GP} here and Fig.~3 of \citealt{sb2014}).
This is because the gas (potential) of the same mass $2M_\ast$ is much more spread out over the
length scale $R_0>R_\ast$ in the present case than in \citet{sb2014} where the gas follows
the highly compact pre-gas-release cluster profile (see above). Hence, the latter is subjected
to significantly more central mass loss than the present pre-gas-release clusters.
The above discussion implies that the combination of the physically motivated initial conditions adopted
here optimally reproduces the observed mass density profile of HD97950.  

\section{Conclusions, discussions and outlook}\label{discuss}

The primary objective of the calculations described in the previous sections
is to find initial conditions under which it is feasible to assemble a VYMC like the HD97950
from an initial distribution of subclusters, given the mass and age constraints
for this cluster. The key inferences can be summarized as follows:

\begin{itemize}

\item A system of subclusters of total stellar mass $M_\ast\approx10^4\Ms$ assemble into a (near) spherical core-halo
star cluster by the age of HD97950 (\ie, in $t<1$ Myr) provided these subclusters are largely
born over a region of scale length more compact than $R_0\lesssim2$ pc. This can happen, \eg, in an intense starburst
event at a dense ``spot'' in a molecular cloud.

\item The initial
sizes of the subclusters are constrained by the highly compact sections of molecular gas filaments or filament
junctions which, in turn, determines the compactness of the final assembled cluster. The size of the final cluster is also
independent of the presence of residual molecular gas. Therefore, the mass density
over the central region (within a virial radius) of the merged cluster is determined by the total stellar
mass that is involved in its assembly.

\item A ``dry'' merger of subclusters, \ie, infall in absence of residual molecular gas (all gas consumed into stars)
always leads to a star cluster that is centrally overdense w.r.t. HD97950, even for the latter's lower mass
limit $M_\ast\approx10^4\Ms$. This is irrespective of the initial mode of subclustering. The expansion
of the final cluster due to stellar winds and primordial binaries is insufficient, as indicated by
the present calculations.

\item A substantial residual gas expulsion ($\approx70$\%) occurring after the formation of the merged cluster
expands the latter to obtain a cluster profile
that is consistent with the observed HD97950. With the lower stellar mass limit $M_\ast\approx10^4\Ms$
and an SFE of $\epsilon\approx30$\%, the observed surface mass density profile of HD97950
can be fairly and optimally reproduced although the central velocity dispersion falls short by $\approx1{\rm~km~s}^{-1}$.

\item In principle, with more extended initial subclusters it is possible to obtain a merged profile that
agrees with that observed for HD97950, without any gas dispersal (\ie, via dry mergers). However,
the resulting central velocity dispersion would then be even lower than the observed value. Much extended
subclusters are as well inconsistent with the highly compact widths of molecular gas filaments.

\item Hence, \emph{the NGC 3603 young cluster (HD97950) has formed essentially monolithically
followed by a substantial gas dispersal.
The initial monolithic stellar distribution has either formed in situ or has been assembled
promptly (in $\lesssim1$ Myr) from closely packed (within $\lesssim2$ pc) less massive stellar clusters (subclusters).
Both scenarios are consistent with the formation of HD97950's entire stellar population in a single starburst
of very small duration.}
The in-situ scenario \citep{sb2014} seems to better reproduce the observed properties of HD97950.   
 
\end{itemize}

A key concern in the above conclusion is that such closely packed gas-embedded
subclusters are \emph{usually} not seen
in practice. This further supports the in-situ formation of young massive clusters.    
However, this does not rule out the prompt assembly channel either.
The initial closely packed subcluster distribution has a very short lifetime
and it merges to a single (embedded) core-halo structure in a fraction of a Myr.
Hence such systems should be rare to observe.
Notably, recent multi-wavelength observations of the Pismis 24 cluster of NGC 6357
\citep{massi2014} indicate that this young cluster (age 1-3 Myr) contains distinct substructures
which must have formed out of dense gas clumps packed within $\approx1$ pc radius.
Using the MYStIX
survey catalog, \citet{jah2014} also find that the stellar distribution in
young clusters (1-3 Myr) tend to smoothen out with age and local stellar density.
This indicates an appearance of these systems as closely packed stellar
overdensities which disappear on a dynamical timescale as seen in the
calculations described in the above sections.

Also, the requirement of substantial ($\approx70$\% by mass) gas expulsion
is supported by the lack of gas in young and intermediate-aged clusters. In particular,
a recent survey of the LMC's massive star clusters over wide ranges of mass ($>10^4\Ms$) and age (30-300 Myr) has failed
to identify reserved gas in any of these clusters \citep{bs2014}.
These clusters would have accreted enough surrounding gas
by now for the latter to be detected inside them. This implies that star clusters can, in fact,
disperse their gaseous component efficiently at any age $<300$ Myr and irrespective of their escape velocities
\citep{bs2014}.

The primary simplification in the above study is the analytic treatment of the residual gas that forms
a spherically symmetric smooth Plummer profile (see Sec.~\ref{initgas}).
The gas consumption by the star formation process
in the vicinity of the individual subclusters would imprint a ``swiss-cake'' structure
in the spatial distribution of the density of the residual gas.
Furthermore, the residual gas cloud could still develop filamentary overdensity structures across it.
Hence an infalling subcluster is susceptible to tidal shocks that accelerate its
disruption. This detail is not critical for the present conclusions since the tidal field due to the background
smoothed gas potential already disrupts the subclusters significantly during their infall,
as seen in the above calculations.

It is rather surprising that despite the wide variety of initial conditions (\cf Table~\ref{tab:kfit2})
and the simplistic treatment of gas expulsion, the density profile of HD97950 is
reproduced naturally in each case. This is as well true for the in-situ model
of \citet{sb2014}. As explained there, this implies that the analytic formulation
aptly describes the bulk dispersal of the residual gas. The complex details of
matter-radiation interaction and localized processes (\eg, turbulence, instabilities and
outflows) seemingly do not affect the release of the bulk of the gas from the system.
That the dynamic expulsion of gas, as modelled using the SPH method, reproduces the here-used
analytical time variation has indeed been demonstrated by \citet{gb2001}. 

The use of initially spherical subclusters (with or without the gas potential) is also
an idealization since a newly hatched group of stars would preferentially follow its
local gas overdensity pattern. However, the stellar assembly would become
spherical (ellipsoidal in presence of gas potential) in a few of its dynamical times. 
For a stellar assembly as compact as the gas filaments, its internal dynamical
time is typically much shorter than that of its infall. This effectively reduces a more realistic
system to the idealized initial condition adopted here. For a sufficiently small initial span (\eg, for the smallest $R_0$
used here; \cf panels 2 \& 3 of Fig.~\ref{fig:snaps_0myr}),
the two timescales are comparable and the whole system can be considered to approach
spherical symmetry (and dynamical equilibrium) in a single dynamical timescale. In that case
the geometry of the individual subclusters does not play a role.

It is also possible that the individual subclusters collect a substantial fraction of the residual
molecular gas during their migration towards the global potential minimum. Additionally, gas can be
streamed towards the global potential minimum. Such cooling flows become efficient if the gas is
sufficiently cold. In that event the assembled system will have the
gas following the stellar distribution closely unlike the present case (see Sec.~\ref{mrgexp}).
Hence the system reduces to a monolithic ``initial'' condition as in \citet{sb2013,sb2014} for
which there exists a solution that matches well with HD97950 \citep{sb2014}.

In the present study, $t\approx1$ Myr is considered as a ``deadline'' for cluster formation (see above) since
the focus here is on the formation of HD97950. The zero age is the beginning of the dynamical
evolution (or of the N-body calculations), which corresponds to the epoch of the formation of the subclusters
in a single starburst. The concurrent appearance of subclusters, \ie, a single starburst of very small or
zero age spread giving rise to a subclustered configuration, is plausible only in a sufficiently compact
region. The more extended the star-bursting region is (\ie, the larger is $R_0$), a correlation is necessary over larger
distances which makes such an event increasingly less probable. Hence, the prompt assembly or the compact in-situ
formation are consistent with both the inferred young age and the very small age spread of HD97950. Note
that $t\approx1$ Myr does not represent any fundamental deadline; it is used here to narrow
down the initial conditions.

The present calculations imply that although VYMCs can form as highly substructured
(or fractal) stellar distributions, they still undergo a much more compact (and hence dense)
phase than their present-day state. Hence, an initial primordial binary population would be
substantially more dynamically processed \citep{mk2012}
than what has been suggested recently by \citet{parker2014}, since
the computed systems in the latter study do not necessarily go through a dense enough phase.

A straightforward leap in the present line of study is to conduct a more thorough survey of the
merger time as a function of $M_\ast$, $R_0$ and $\epsilon$, irrespective of
any deadline. It is as well necessary to quantitatively relate the structure of the final cluster
with these parameters and also with the sizes of the initial subclusters. Such a survey
would be applicable to VYMCs in general. The detail of such modelling can be improved in
future by introducing PMS evolution recipes and primordial binaries.

Another important improvement would be to do a detailed structural analysis
of the final merged system. In particular, the construction of adaptive surface
density maps, as in \citet{kuhn2014}, can be applied to the computed merged systems.
This would determine their sphericity accurately or reveal if they still contain
substructures at a given age. By comparing with the density patterns of observed VYMCs as in \citet{kuhn2014},
it would be possible to constrain the initial conditions of these VYMCs. In this
context, it would be worthwhile to obtain such density contours for the observed HD97950 cluster as well
in order to determine whether it harbours substructures. This would further constrain
the conditions at its birth.

It is currently technologically formidable to form the subclusters from
ab-initio hydrodynamic calculations for the present mass scale ($>10^4\Ms$).
However, in the foreseeable future, it would be possible to treat the residual
gas hydrodynamically that can be energized by an appropriately modelled stellar
feedback. This is possible to achieve by, \eg, using the ``AMUSE'' framework \citep{pz2008}.
In this way the gas expulsion process can be treated in a more realistic but
technologically accessible manner (Simon Portegies Zwart; private communications).
The timescale of the dispersal would, of course, depend on the
modelling of the stellar feedback.

The above discussions imply that although simplifications and idealizations are
adopted in the present study, they do not alter the key conclusions and their
interrelations as enlisted above, namely, that the evidence points
towards VYMCs such as HD97950, R136, the ONC and the Pleiades having
formed essentially monolithically with substantial gas expulsion.
Future unprecedented resolution of molecular cloud filaments and filament
junctions with \emph{ALMA} and proper motion measurements of the stars in the outer regions
of young star clusters with \emph{Gaia} (see \citealt{sb2014}) will provide more direct tests
of such birth environments of VYMCs.

\section*{Acknowledgements}
The authors are thankful to 
Pau Amaro-Seoane of the Albert Einstein Institute, Potsdam, 
Nate Bastian of the Liverpool John Moores University,
Roberto Capuzzo-Dolcetta of the La Sapienza University, Rome,
Mike Fellhauer of the University of Concepcion and
Adam Grinsberg of the European Southern Observatory, Garching for useful comments and criticisms.
The authors are thankful to the anonymous referee for the suggestions
which have improved the presentation of the paper.

\newpage

\begin{table*}
\begin{minipage}{7.0 in}
\renewcommand{\thefootnote}{\alph{footnote}}
\caption{Parameters of the NGC 3603 Young Cluster (HD 97950) as determined from
observations.}
\label{tab:hd9795}
\begin{tabular}{llll}
\hline
Quantity & Measurement & Value(s) & Reference(s)\\
\hline
Mass $M_\ast$  & Photometry & $(1.0-1.6)\times10^4\Ms$ & \citet{stol2006,hara2008}\\
Mass $M_\ast$  & Kinematics & $(1.7-1.9)\times10^4\Ms$ & \citet{roch2010,pang2013}\\
Age & Photometry & 1 Myr & \citet{stol2004,pang2013}\\
Velocity dispersion $\sigma_{\rm 1d}$ (mean 1-d) & 
                    Proper motion & $4.5\pm0.8{\rm~km~s}^{-1}$ & \citet{roch2010}\\
Velocity dispersion $\sigma_{\rm 1d}$(orthogonal 1-d) & 
                    Proper motion & $(4.8-6.5)\pm0.5{\rm~km~s}^{-1}$ & \citet{pang2013}\\ 
Central concentration parameter $k$ & King/EFF profile fit
                                      \footnote{The theoretical King (also Elson-Fall-Freeman or EFF)
profiles are best-fitted with the observed surface mass density profile of HD 97950 for
the stellar mass range $0.5-2.5\Ms$. See also Table~\ref{tab:harafit}.}
                                      & 4.2 &  \citet{hara2008}\\ 
Core radius $r_c$ &  King/EFF profile fit & $4.8''(\approx0.15{\rm~pc})$ & \citet{hara2008}\\
Tidal radius $r_t$ & King/EFF profile fit & large &  \citet{hara2008}\\
Distance from the Sun & Photometry & $6.0\pm0.8(0.3)$ kpc 
                                              & \citet{hara2008} (\citealt{stol2004})\\    
\hline 
\end{tabular}
\end{minipage}
\end{table*}

\begin{table*}
\begin{minipage}{4.5 in}
\caption{A basic classification of the different morphologies in the spatial distribution
of stars that can occur in the process of subcluster merging. These morphologies appear
in the models computed here. Note that the distinctions among these morhphologies
are only qualitative for the present purpose and are made for the ease of descriptions.}
\label{tab:morphdef}
\begin{tabular}{lcl}
\hline
Morphology & Abbreviation & Example\\
\hline 
Substructured                               & SUB               & Fig.~\ref{fig:snaps_0myr}\\
Core + asymmetric and/or substructured halo & CHas              & Fig.~\ref{fig:snaps83_1myr}, panels 3,4\\
Core-halo with near spherical symmetry      & CH                & Fig.~\ref{fig:snaps_egevol}, panel 3\\
Core + halo containing satellite clusters   & CHsat             & Fig.~\ref{fig:snaps7_R5}, panel 4\\
\hline
\end{tabular}
\end{minipage}
\end{table*}

\begin{table*}
\begin{minipage}{6.5 in}
\renewcommand{\thefootnote}{\alph{footnote}}
\caption{An overview of the evolutionary sequences of the primary systems as computed here. A particular row
indicates how the morphology (see Table~\ref{tab:morphdef})
of the corresponding system evolves with evolutionary time (in Myr as indicated by the numerical
values along the columns 3-5 and 6-8), for systems both without and with a background gas potential (see text).
As expected, the systems, in general, evolve from substructured to a core-halo configuration with a timescale
that increases with increasing initial extent $R_0$. See text for details.}
\label{tab:morphevol}
\begin{tabular}{cclllclll}
\hline
Config. name & Short name
& \multicolumn{3}{c}{Without gas potential} &  &
\multicolumn{3}{c}{With gas potential $(\epsilon\approx0.3)$}\\
\hline
m1000r0.1R1.1N10            
\footnote{m$x$r$y$R$z$N$n$ implies an initial system (at $t=0$)
comprising of N$=n$ Plummer clusters, each of mass m$=m_{cl}(0)=x\Ms$
and half-mass radius r$=r_h(0)=y$ pc, distributed uniformly over a spherical volume of radius R$=R_0=z$ pc.}
&     A-Ia     &        0.2,SUB&0.6,CHas&1.0,CH       &  &     0.2,SUB&0.6,CHas&1.0,CH\\
m1000r0.3R1.1N10            &     A-Ib     &        0.2,SUB&0.6,CHas&1.0,CHas     &  &     0.2,SUB&0.6,CHas&1.0,CHas\\
m1000r0.1R2.5N10            &     A-IIa	   &        0.6,SUB&1.0,SUB&2.0,CH        &  &     0.6,SUB&1.0,CHsat&2.0,CHas\\
m1000r0.3R2.5N10            &     A-IIb    &        0.6,SUB&1.0,SUB&2.0,CHas      &  &     0.6,SUB&1.0,CHsat&2.0,CHas\\
m10-150r0.01-0.1R1.1N150    
\footnote{Further, when a range of values $x1-x2$ is used instead of a single value, it implies that the corresponding
quantity is uniformly distributed over $\left[x1,x2\right]$ at $t=0$.}
&     B-Ic     &        0.2,SUB&0.6,CHas&1.0,CH       &  &     0.2,SUB&0.6,CHas&1.0,CHas\\
m10-150r0.1R1.1N150         &     B-Ia     &        0.2,SUB&0.6,CHas&1.0,CH       &  &     0.2,SUB&0.6,CHas&1.0,CHas\\
m10-150r0.1R2.5N150         &     B-IIa    &        0.6,SUB&1.0,SUB&2.0,CHas      &  &     0.6,SUB&1.0,CHas&2.0,CHas\\
m1000r0.5-1.0R5.0N10        &     A-IIId   &        1.0,SUB&2.0,SUB&3.0,CHas      &  &     1.0,CHsat&2.0,CHsat&3.0,CHsat\\
m1000r0.5-1.0R10.0N10       &     A-IVd    &        1.0,SUB&2.0,SUB&3.0,SUB       &  &     1.0,SUB&2.0,SUB&3.0,CHsat\\
\hline
\end{tabular}
\end{minipage}
\end{table*}

\begin{table*}
\begin{minipage}{3.2 in}
\renewcommand{\thefootnote}{\alph{footnote}}
\caption{Best-fit parameters for the King surface mass-density profile with
that observed for the NGC 3603 young cluster \citep{hara2008}.}
\label{tab:harafit}
\begin{tabular}{cllll}
\hline
Data included & $k$ & $r_c$($''$) & $r_t$($''$) & $\chi^2$\\
\hline
All                  &   4.3$\pm$0.6   &  4.8$\pm$0.4      &     --- \footnotemark[2]    &     1.5\\
All but 2nd annule\footnotemark[1]
                     &   3.8$\pm$0.6   &  5.1$\pm$0.4      &     --- \footnotemark[2]    &     1.2\\
All but 1st annule\footnotemark[1]
                     &   5.8$\pm$1.1   &  4.1$\pm$0.4      &     --- \footnotemark[2]    &     1.2\\
\hline
\footnotetext[1]{The measured projected mass-density profile of \citet{hara2008}
shows substantial fluctuations among the innermost annuli. Equally good but fitted profiles
of significantly different central concentrations ($k$) are obtained when each of the
two innermost annuli are considered individually. At the distance of NGC 3603 $1''\approx0.03$ pc.}
\footnotetext[2]{The King tidal radius ($r_t$) is large and does not affect the fit, implying a
cluster profile with no tidal cut-off.}
\end{tabular}
\end{minipage}
\end{table*}

\begin{table*}
\begin{minipage}{4.5 in}
\renewcommand{\thefootnote}{\alph{footnote}}
\caption{Best-fit parameters for the King surface mass-density profile at $t\approx1$ Myr for those
computed configurations which evolve to form a star cluster with near-spherical core-halo structure
(the CH-type morphology, see Table~\ref{tab:morphdef}) within $t<1$ Myr, in absence of a
background gas potential. All of these profiles are much more concentrated and/or flattened
than the observed profile of HD97950 (\cf, Table~\ref{tab:harafit}) as is also evident from
Fig.~\ref{fig:dryprofs_1myr}.}
\label{tab:kfit1}
\begin{tabular}{cclllll}
\hline
Config. name & Short name
& $k$ & $r_c$($''$) & $r_t$($''$)
& $\chi^2$ & $R_{\rm lim}$($''$)\footnotemark[1]\\
\hline
m1000r0.1R0.5N10            &     A-0.5Ia    &       26.7   &   4.4   &   37.0  &   1.3  &    25\\
m1000r0.1R1.1N10            &     A-Ia       &        8.2   &   7.1   &   88.2  &   0.5  &    40\\
m1000r0.1R2.5N10            &     A-IIa      &      --- & --- & --- & --- & ---\footnotemark[2]\\
\hline
m1000r0.3R0.5N10            &     A-0.5Ib    &       16.6   &   3.7   &  126.1  &   1.4  &    65\\
m1000r0.3R1.1N10            &     A-Ib       &        9.6   &   6.2   &   87.7  &   1.8  &    40\\
m1000r0.3R1.5N10            &     A-1.5Ib    &       10.3   &   7.2   &   57.7  &   1.1  &    40\\
m1000r0.3R2.0N10            &     A-2.0Ib    &        5.5   &   6.8   &  254.1  &   1.0  &    40\\
m1000r0.3R2.5N10            &     A-IIb      &      --- & --- & --- & ---& ---\\
\hline
m10-150r0.01-0.1R1.1N150    &     B-Ic       &        9.3   &   7.2   &   85.3  &   1.7  &    65\\
m10-150r0.1R1.1N150         &     B-Ia       &       11.0   &   5.6   &   83.3  &   2.3  &    50\\
m10-150r0.1R2.5N150         &     B-IIa      &      --- & --- & --- & ---& ---\\
\hline
m1000r0.3R0.5N10-se\footnotemark[3] 
                            &     A-0.5Ib-se &       17.7   &   3.5   &  137.0  &   2.2  &    65\\
m1000r0.3R1.1N10-pb\footnotemark[3]
                            &     A-Ib-pb    &        7.0   &  10.8   &   46.9  &   2.5  &    35\\
m10-150r0.1R1.1N150-se      &     B-Ia-se    &       10.3   &   6.2   &   77.0  &   2.0  &    55\\
\hline
\end{tabular}
\footnotetext[1]{$R_{\rm lim}$ is the distance from the cluster's (density) center until which
a reasonable King-profile fit to its surface density profile could be obtained.}
\footnotetext[2]{The empty data lines indicate that a star cluster with a CH-type morphology cannot
form within $t<1$ Myr.}
\footnotetext[3]{The suffixes ``-se'' and ``-pb'' indicate calculations that include
stellar evolution and a population of primordial binaries (see Sec.~\ref{mrgcls}) respectively.}
\end{minipage}
\end{table*}

\begin{table*}
\begin{minipage}{5.1 in}
\caption{Best-fit parameters for the King surface mass-density profile at $t\approx1$ Myr for
computed post-gas-expulsion configurations. Here, the systems evolve in a background residual gas potential
(see Secs.~\ref{initgas} \& \ref{mrgexp}) to form
a star cluster with near-spherical core-halo structure (the CH-type morphology, see Table~\ref{tab:morphdef})
within $t<1$ Myr followed by the expulsion of the residual gas at $\tau_d\approx0.6$ Myr. The resulting cluster
profiles agree well with the observed projected mass-density profile of HD97950 (\cf, Table~\ref{tab:harafit},
Fig.~\ref{fig:GPprofs_1myr}). All the legends are the same as in Table~\ref{tab:kfit1}.}
\label{tab:kfit2}
\begin{tabular}{cclllll}
\hline
Config. name & Short name & $k$ & $r_c$($''$) & $r_t$($''$) & $\chi^2$ & $R_{\rm lim}$($''$)\\
\hline
m1000r0.1R1.1N10     &            A-Ia     &          2.3$\pm$0.1  &  9.5$\pm$0.7 &  672.4$\pm$712  & 0.9    &    45.0\\
                     &                     &          2.4$\pm$0.2  &  9.0$\pm$0.5 & 1242.6$\pm$626  & 1.1    &    110.0\\
                     &                     &                       &              &                 &        &       \\
m1000r0.3R1.1N10     &            A-Ib     &          5.7$\pm$0.6  &  5.8$\pm$0.6 &  269.3$\pm$180  &  2.4   &     40.0\\
                     &                     &          6.0$\pm$0.7  &  5.1$\pm$0.4 & 2110.7$\pm$2574 &  2.6   &     110.0\\
                     &                     &                       &              &                 &        &       \\
m10-150r0.1R1.1N150  &            B-Ia     &          5.4$\pm$0.5  &  6.5$\pm$0.6 &  160.6$\pm$44   & 2.6    &     50.0\\
                     &                     &          5.6$\pm$0.6  &  5.5$\pm$0.4 &  697.9$\pm$284  & 3.1    &     110.0\\
                     &                     &                       &              &                 &        &       \\
m1000r0.3R1.1N10-pb  &            A-Ib-pb  &          4.2$\pm$0.8  &  6.8$\pm$0.8 &  934.1$\pm$435  & 1.6    &   $6.0-110.0$\\ 
                     &                     &          2.6$\pm$0.2  & 10.8$\pm$1.0 &  174.5$\pm$43   & 2.4    &   $55.0$\\
\hline
\end{tabular}

\medskip
The quality of the King profile fit degrades moderately (see the $\chi^2$ column)
if the outer halos of the clusters are included ($R_{\rm lim}\approx110''$)
when compared to that for only their inner regions ($40''\lesssim R_{\rm lim}\lesssim 60''$). In both cases, the fits are
consistent with large values of $r_t$ implying untruncated, monolithic cluster profiles as observed in
HD97950 (see Table~\ref{tab:harafit}).
\end{minipage}
\end{table*}

\label{lastpage}

\end{document}